\documentclass[acmsmall]{acmart}

\renewcommand\footnotetextcopyrightpermission[1]{}

\usepackage{subfiles}
\usepackage{preamble}

\AtBeginDocument{%
  }

\begin{document}


\newcommand{\locspec}{28,862}

\newcommand{\nSyn}{591}
\newcommand{\nRel}{253}
\newcommand{\nRule}{1,090}
\newcommand{\nDec}{561}


\newcommand{\POS}{{\small\texttt{Pos}}}
\newcommand{\NEG}{{\small\texttt{Neg}}}

\newcommand{\POSSmall}{\footnotesize\texttt{Pos}}
\newcommand{\NEGSmall}{\footnotesize\texttt{Neg}}

\newcommand{\nPOS}{1,307}
\newcommand{\nNEGTbl}{\phantom{0,}564}
\newcommand{\nNEG}{564}

\newcommand{\nPOSPass}{1,216}
\newcommand{\nNEGPassTbl}{\phantom{0,}521}
\newcommand{\nNEGPass}{521}

\newcommand{\nPOSFail}{0}
\newcommand{\nNEGFail}{0}

\newcommand{\nPOSExcl}{91}
\newcommand{\nNEGExcl}{43}
\newcommand{\nTYPEExcl}{134}



\newcommand{\nTYPEExclFuture}{18}
\newcommand{\nTYPEExclSpec}{24}
\newcommand{\nTYPEExclComp}{30}
\newcommand{\nTYPEExclSpecific}{62}



\newcommand{\VMOD}{{\small\texttt{V1Model}\xspace}}
\newcommand{\EBPF}{{\small\texttt{eBPF}\xspace}}
\newcommand{\PSA}{{\small\texttt{PSA}\xspace}}

\newcommand{\VMODSmall}{\footnotesize\texttt{V1Model}\xspace}
\newcommand{\EBPFSmall}{\footnotesize\texttt{eBPF}\xspace}
\newcommand{\PSASmall}{\footnotesize\texttt{PSA}\xspace}

\newcommand{\VMODPlus}{{\small\texttt{V1Model+}\xspace}}
\newcommand{\EBPFPlus}{{\small\texttt{eBPF+}\xspace}}

\newcommand{\VMODPlusSmall}{\footnotesize\texttt{V1Model+}\xspace}
\newcommand{\EBPFPlusSmall}{\footnotesize\texttt{eBPF+}\xspace}

\newcommand{\nVMODTbl}{\phantom{0,}203}
\newcommand{\nVMOD}{203}
\newcommand{\nVMODPassTbl}{\phantom{0,}194}
\newcommand{\nVMODPass}{194}
\newcommand{\nVMODPatch}{0}
\newcommand{\nVMODFail}{0}
\newcommand{\nVMODExclStTbl}{\phantom{00}6}
\newcommand{\nVMODExclSt}{6}
\newcommand{\nVMODExclDyTbl}{\phantom{0}3}
\newcommand{\nVMODExclDy}{3}

\newcommand{\nVMODPlus}{1,982}
\newcommand{\nVMODPlusPassTbl}{\phantom{0}1,841}
\newcommand{\nVMODPlusPass}{1,841}
\newcommand{\nVMODPlusPatch}{2}
\newcommand{\nVMODPlusFail}{0}
\newcommand{\nVMODPlusExclSt}{102}
\newcommand{\nVMODPlusExclDy}{39}

\newcommand{\nEBPFTbl}{\phantom{0,0}17}
\newcommand{\nEBPF}{17}
\newcommand{\nEBPFPassTbl}{\phantom{0,0}15}
\newcommand{\nEBPFPass}{15}
\newcommand{\nEBPFPatch}{0}
\newcommand{\nEBPFFail}{0}
\newcommand{\nEBPFExclStTbl}{\phantom{00}0}
\newcommand{\nEBPFExclSt}{0}
\newcommand{\nEBPFExclDyTbl}{\phantom{0}2}
\newcommand{\nEBPFExclDy}{2}

\newcommand{\nEBPFPlusTbl}{\phantom{0,}144}
\newcommand{\nEBPFPlus}{144}
\newcommand{\nEBPFPlusPassTbl}{\phantom{0,}133}
\newcommand{\nEBPFPlusPass}{133}
\newcommand{\nEBPFPlusPatch}{0}
\newcommand{\nEBPFPlusFail}{0}
\newcommand{\nEBPFPlusExclStTbl}{\phantom{00}0}
\newcommand{\nEBPFPlusExclSt}{0}
\newcommand{\nEBPFPlusExclDy}{11}

\newcommand{\nPSATbl}{\phantom{0,0}26}
\newcommand{\nPSA}{26}
\newcommand{\nPSAPassTbl}{\phantom{0,0}24}
\newcommand{\nPSAPass}{24}
\newcommand{\nPSAPatch}{0}
\newcommand{\nPSAFail}{0}
\newcommand{\nPSAExclStTbl}{\phantom{00}1}
\newcommand{\nPSAExclSt}{1}
\newcommand{\nPSAExclDyTbl}{\phantom{0}1}
\newcommand{\nPSAExclDy}{1}


\newcommand{\nINTERPExclComp}{10}
\newcommand{\nINTERPExclSpecific}{46}
\newcommand{\nINTERPExcl}{56}
\newcommand{\nTYPEExclNoLimit}{134}



\newcommand{\nELPrem}{3,865}
\newcommand{\nELRulePrem}{1,285}
\newcommand{\nELIfPrem}{2,580}

\newcommand{\nILPrem}{8,304}
\newcommand{\nILRulePrem}{1,285}
\newcommand{\nILIfPrem}{2,519}
\newcommand{\nILLetPrem}{4,500}

\newcommand{\nELtoILDiff}{4,439}

\newcommand{\nSLInstr}{6,282}
\newcommand{\nSLRuleInstr}{\phantom{0,}838}
\newcommand{\nSLIfInstr}{2,265}
\newcommand{\nSLLetInstr}{3,179}

\newcommand{\nSLBranchInjected}{649}


\newcommand{\nPOSCompare}{1,216}
\newcommand{\nPOSPetraPass}{1,086}
\newcommand{\nPOSPetraFail}{130} 
\newcommand{\rPOSPetraPass}{89.3\%}
\newcommand{\nPOSSPECTECPass}{1,216}
\newcommand{\rPOSSPECTECPass}{100.0\%}

\newcommand{\nNEGCompareTbl}{\phantom{0,}521}
\newcommand{\nNEGCompare}{521}
\newcommand{\nNEGPetraPassTbl}{\phantom{0,}477}
\newcommand{\nNEGPetraPass}{477}
\newcommand{\nNEGPetraFail}{44} 
\newcommand{\rNEGPetraPass}{91.6\%}
\newcommand{\nNEGSPECTECPassTbl}{\phantom{0,}521}
\newcommand{\nNEGSPECTECPass}{521}
\newcommand{\rNEGSPECTECPass}{100.0\%}

\newcommand{\nVMODCompareTbl}{\phantom{0,}194}
\newcommand{\nVMODCompare}{194}
\newcommand{\nVMODPetraPassTbl}{\phantom{0,}187}
\newcommand{\nVMODPetraPass}{187}
\newcommand{\rVMODPetraPass}{96.4\%}
\newcommand{\nVMODHOLPassTbl}{\phantom{0}58}
\newcommand{\nVMODHOLPass}{58}
\newcommand{\rVMODHOLPass}{29.9\%}
\newcommand{\nVMODSPECTECPassTbl}{\phantom{0,}194}
\newcommand{\nVMODSPECTECPass}{194}
\newcommand{\rVMODSPECTECPass}{100.0\%}

\newcommand{\nVMODPlusCompare}{1,841}
\newcommand{\nVMODPlusPetraPass}{1,120}
\newcommand{\rVMODPlusPetraPass}{60.8\%}
\newcommand{\nVMODPlusHOLPass}{355}
\newcommand{\rVMODPlusHOLPass}{19.3\%}
\newcommand{\nVMODPlusSPECTECPass}{1,839}
\newcommand{\rVMODPlusSPECTECPass}{99.9\%}

\newcommand{\nEBPFCompareTbl}{\phantom{0,0}15}
\newcommand{\nEBPFCompare}{15}
\newcommand{\nEBPFPetraPassTbl}{\phantom{0,00}6}
\newcommand{\nEBPFPetraPass}{6}
\newcommand{\rEBPFPetraPass}{40.0\%}
\newcommand{\nEBPFHOLPassTbl}{\phantom{00}4}
\newcommand{\nEBPFHOLPass}{4}
\newcommand{\rEBPFHOLPass}{26.7\%}
\newcommand{\nEBPFSPECTECPassTbl}{\phantom{0,0}15}
\newcommand{\nEBPFSPECTECPass}{15}
\newcommand{\rEBPFSPECTECPass}{100.0\%}

\newcommand{\nEBPFPlusCompareTbl}{\phantom{0,}133}
\newcommand{\nEBPFPlusCompare}{133}
\newcommand{\nEBPFPlusPetraPassTbl}{\phantom{0,}114}
\newcommand{\nEBPFPlusPetraPass}{114}
\newcommand{\rEBPFPlusPetraPass}{85.7\%}
\newcommand{\nEBPFPlusHOLPassTbl}{\phantom{0}86}
\newcommand{\nEBPFPlusHOLPass}{86}
\newcommand{\rEBPFPlusHOLPass}{64.7\%}
\newcommand{\nEBPFPlusSPECTECPassTbl}{\phantom{0,}133}
\newcommand{\nEBPFPlusSPECTECPass}{133}
\newcommand{\rEBPFPlusSPECTECPass}{100.0\%}


\newcommand{\SP}[1]{{\small\textbf{SP#1}}}
\newcommand{\CB}[1]{{\small\textbf{CB#1}}}

\newcommand{\nSpecIssue}{15}
\newcommand{\nCompIssue}{20}

\newcommand{\nSpecPatch}{9}
\newcommand{\nCompBug}{15}
\newcommand{\nBug}{24} 

\newcommand{\nSPSevenDep}{98}

\title{P4-SpecTec: Integrating a Language Mechanization Framework into the Real-World P4 Specification}


\author{Jaehyun Lee}
\email{99jaehyunlee@kaist.ac.kr}
\affiliation{%
  \institution{KAIST}
  \country{South Korea}
}

\author{Seokhun Jeong}
\email{sraccoon@kaist.ac.kr}
\affiliation{%
  \institution{KAIST}
  \country{South Korea}
}

\author{Sehyuk Ahn}
\email{shyukahn@kaist.ac.kr}
\affiliation{%
  \institution{KAIST}
  \country{South Korea}
}

\author{Haechan Kwon}
\email{pacokwon@kaist.ac.kr}
\affiliation{%
  \institution{KAIST}
  \country{South Korea}
}

\author{Sukyoung Ryu}
\email{sryu.cs@kaist.ac.kr}
\affiliation{%
  \institution{KAIST}
  \country{South Korea}
}

\begin{abstract}

  Programming languages evolve over time, but often without a
  complete and unambiguous definition of their syntax and semantics.
  Ambiguities and inconsistencies are silently introduced into specifications,
  and manifest as divergences between the specification, implementations, and
  formalizations that constitute the language ecosystem.
  Even in rare cases when a normative specification exists, like JavaScript and
  WebAssembly (Wasm), keeping the ecosystem in sync is a daunting task.
  Language mechanization frameworks address this problem by treating a
  mechanized specification as the \emph{single source of truth}, from which
  implementations and documents are generated.
  Recently, this approach has been integrated into the actual JavaScript
  and Wasm specifications with ESMeta and Wasm-SpecTec, respectively.
  Despite these successes, it remains an open question how to extrapolate
  ESMeta and Wasm-SpecTec to other language specifications.
  Both framework designs leverage the existence of JavaScript and Wasm's
  normative specifications, which is not the case for many languages.

  As a first step towards addressing this question, we present P4-SpecTec, a
  language mechanization framework for the P4 programming language, as a case
  study of real-world adoption of language mechanization.
  P4 is a statically-typed domain-specific language for programming packet
  processors.
  It is evolving without a normative specification, thereby introducing
  inconsistencies and errors into the P4 ecosystem.
  From a mechanization framework perspective, P4 introduces unique challenges,
  in particular the requirement that its type system mechanization should be
  executable, which is not supported by either ESMeta or Wasm-SpecTec.
  To address this challenge, we introduce \emph{algorithmic inference rules} as
  the primary instrument for mechanization, enabling the mechanized P4 static
  and dynamic semantics to be executed as a P4 type checker and interpreter,
  respectively.
  We mechanized the most recent P4 specification, and utilizing its
  executability, identified \nBug{} bugs across the official P4 specification
  and the reference compiler.
  Furthermore, P4-SpecTec derives a specification document as prose algorithms,
  making it accessible to P4 developers.
  P4-SpecTec is conditionally adopted as the official P4 specification authoring
  toolchain.
  We share the lessons learned from our case study, to provide insights for
  integrating mechanization into real-world languages without normative
  specifications.

\end{abstract}

\maketitle

\pagestyle{plain}
\thispagestyle{plain}

\section{Introduction}\label{sec:introduction}

Modern programming language specifications evolve.
New features are added to match the growing needs of developers and emerging
hardware capabilities.
Additionally, specifications are updated as they are not free of bugs.
Because language semantics is inherently complex, inconsistencies and
ambiguities within the specification are routinely discovered and corrected.
Such evolution occurs in the presence of multiple language representations.
A language is defined across multiple artifacts, including the specification,
reference implementations, formalizations, and test suites.
These provide different views of the same language, each serving a distinct
purpose.
While evolving, they must remain synchronized, coherently describing the
intended behavior of the language.

Evolving specifications bring progress but also challenges.
As languages grow, it becomes increasingly difficult to reason about how newly
proposed features interact with existing ones.
Consequently, it becomes more likely that inconsistencies and ambiguities are
introduced without being noticed.
Over time, they manifest as divergent interpretations across different
representations.
This creates confusion about the intended semantics among users and
implementers.
%

Fundamentally, a normative specification that is both complete and unambiguous
is essential to rectify these issues.
However, such a normative specification often exists only in the minds of the
language designers.
The conceptual model is not even monolithic; each designer specializes only in
certain aspects of the language, leading to a fragmented understanding of the
overall language.
In rare cases, languages such as Standard ML~\cite{sml},
JavaScript~\cite{ecmascript}, and WebAssembly (Wasm)~\cite{wasm} have
well-established normative specifications.
In Fig.~\ref{fig:esmeta-and-wasm-spectec}, (a) and (c) illustrate the
JavaScript and Wasm specifications for their respective \pfourinl|if| constructs.
The JavaScript specification defines the language in terms of rigorous prose
algorithms, while the Wasm specification uses formal mathematical definitions.
Despite their rigor, specifications are typically maintained manually in plain
text format.
Hence, achieving consistent evolution of a language still remains a significant
challenge.

\emph{Language mechanization frameworks} address this challenge by providing
the means to mechanize specifications.
Mechanize, in this context, means to define the specification using a
machine-readable \emph{meta-language}.
%
%
Specifications can then be parsed, executed, or translated into
various language representations as backends.
Evolution is then a matter of updating the mechanized specification, with
systematic checks for errors and representation divergences.
Frameworks such as the K Framework~\cite{kframework},
PLT-Redex~\cite{pltredex}, and Ott~\cite{ott} have been developed to support
this approach.

Language mechanization frameworks only partly address the challenges of
evolving specifications.
To complete the picture, mechanization should be integrated into the ecosystem
as the normative specification.
Without integration, mechanization remains separate from the language design,
which must then be maintained and updated in parallel with the original
specification.
Practically, mechanization should serve as the shared vocabulary for language
design discussions, and the official model for design decisions.
Thus, it presents an open question:

\vspace{0.1em}
\begin{center}
\emph{How should mechanization be integrated into real-world language specifications?}
\end{center}
\vspace{0.1em}

\noindent
In this paper, we present a concrete case study on this question and share the lessons
learned.

\subsection{Language Mechanization for Real-World Specifications}\label{sec:intro-mech}

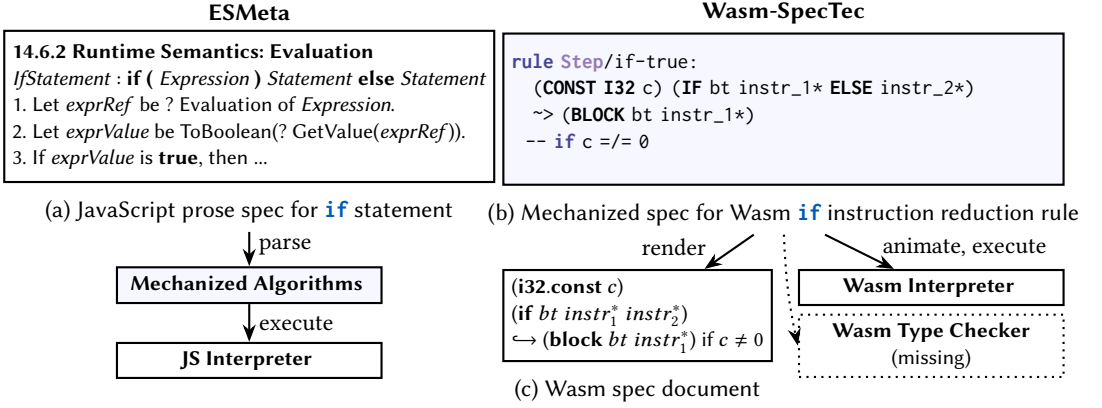
\begin{figure}[t]
  \centering
  \begin{tikzpicture}[
  node distance=1cm and 2cm,
  box/.style={
    draw, thick, align=left,
    font=\sffamily\footnotesize, fill=white,
    inner sep=3pt, minimum width=3.5cm
  },
  title/.style={font=\sffamily\bfseries\small},
  arrow/.style={-{Stealth[scale=1.0]}, thick}
]


\node[title] (es_label) {ESMeta};
\node[box, below=0cm of es_label, minimum height=2cm] (es_prose) {
  \textbf{14.6.2 Runtime Semantics: Evaluation} \\
  \textit{IfStatement} : \textbf{if (} \textit{Expression} \textbf{)} \textit{Statement} \textbf{else} \textit{Statement} \\
  1. Let \textit{exprRef} be ? Evaluation of \textit{Expression}. \\
  2. Let \textit{exprValue} be ToBoolean(? GetValue(\textit{exprRef})). \\
  3. If \textit{exprValue} is \textbf{true}, then ...
};
\node[below=0.1cm of es_prose, font=\sffamily\small] (es_caption) {
  (a) JavaScript prose spec for \pfourinl|if| statement
};
\node[box, below=1.1cm of es_prose, fill=blue!3] (es_mech) {
  \textbf{Mechanized Algorithms}
};
\draw[arrow] (es_caption) -- node[right, font=\small, yshift=1pt] {parse} (es_mech);
\node[box, below=0.5cm of es_mech] (js_interp) {
  \textbf{JS Interpreter}
};
\draw[arrow] (es_mech) -- node[right, font=\small] {execute} (js_interp);


\node[title, right=5.2cm of es_label] (wasm_label) {Wasm-SpecTec};
\node[box, below=0cm of wasm_label, fill=blue!3, text width=7.2cm, minimum height=2cm] (wasm_mech) {
  \vspace{-0.3cm}
  \lstinputlisting[
    style=spectecstyle,
    language=spectec,
    basicstyle=\ttfamily\footnotesize,
    backgroundcolor=\color{blue!3},
    aboveskip=0pt,
    belowskip=0pt
  ]{resources/intro/wasm-if.watsup}
  \vspace{-0.3cm}
};
\node[below=0.1cm of wasm_mech, font=\sffamily\small] (wasm_mech_caption) {
  (b) Mechanized spec for Wasm \pfourinl|if| instruction reduction rule
};
\node[below=1.1cm of wasm_mech] (wasm_phantom) {};
\node[box, below=1.1cm of wasm_mech.south west, anchor=north west, minimum height=1cm] (wasm_formal) {
  (\textbf{i32}.\textbf{const} $c$) \\
  (\textbf{if} $bt$ $instr_1^*$ $instr_2^*$) \\
  $\hookrightarrow$ (\textbf{block} $bt$ $instr_1^*$) \text{if} $c \neq 0$
};
\node[below=0.1cm of wasm_formal, font=\sffamily\small] (wasm_formal_caption) {
  (c) Wasm spec document
};
\node[box, below=1.1cm of wasm_mech.south east, anchor=north east] (wasm_interp) {
  \textbf{Wasm Interpreter}
};
\node[box, align=center, below=0.1cm of wasm_interp, dotted] (wasm_typecheck) {
  \textbf{Wasm Type Checker} \\ (missing)
};

\draw[arrow] (wasm_mech_caption) -- node[left=5pt, font=\small, yshift=2pt] {render} (wasm_formal);
\draw[arrow] (wasm_mech_caption) -- node[right=5pt, font=\small, yshift=2pt] {animate, execute} (wasm_interp);
\draw[arrow, dotted, thick, bend right=5] (wasm_mech_caption) to (wasm_typecheck.west);

\end{tikzpicture}
  \vspace{-2em}
  \caption{
    (Left) ESMeta operating downstream of the ECMAScript specification, parsing
    and reconstructing it. (Right) Wasm-SpecTec operating upstream, replacing
    the original specification with the mechanized one.}
  \label{fig:esmeta-and-wasm-spectec}
  \vspace{-1em}
\end{figure}

Recently, two frameworks have advanced to the point of integration into
real-world specifications: ESMeta~\cite{esmeta} for JavaScript and
Wasm-SpecTec~\cite{wasmspectec} for Wasm.
ESMeta has been officially integrated into the CI system of the JavaScript
standard and its test suite~\cite{esmetaadopt}.
The Wasm Community Group has adopted Wasm-SpecTec as the official toolchain for
authoring and maintaining the specification~\cite{wasmspectecadopt}.
These demonstrate the feasibility and benefits of mechanization at industry
scale.

However, it remains unanswered how to extrapolate from these cases to other
real-world languages.
A common factor underlies both successes: the existence of a well-established
normative specification.
Both framework designs were guided by their respective specifications, as
illustrated in Fig.~\ref{fig:esmeta-and-wasm-spectec}.
ESMeta, on the left, operates downstream of the original specification.
It parses the prose specification and reconstructs it as a set of mechanized
algorithms.
The mechanized algorithms are then executed, effectively yielding a JavaScript
interpreter conforming to the specification~\cite{jiset}.

Wasm-SpecTec, in contrast, moves mechanization upstream by rewriting the
original specification in (c) with a mechanized one in (b).
It uses a declarative meta-language to define the specification, designed to
closely resemble the ``textbook-style'' formalization of the original
specification.
The mechanization is rendered as the official document, regenerating (c).
The declarative mechanization is also translated, or \emph{animated},
into an algorithmic form to yield a Wasm interpreter~\cite{wasmspectec}.

In short, ESMeta reconstructs, whereas Wasm-SpecTec rewrites the existing
normative specification.
The shape of the original specification is hard-wired into the design of both
frameworks.
Thus, the following refined question still remains unexplored:

\vspace{0.1em}
\begin{center}
\emph{How should mechanization be integrated into real-world languages, lacking
  a normative specification?}
\end{center}
\vspace{0.1em}

\noindent
The question is still too broad to be answered in general for all languages.
In this paper, we thus further refine the question by focusing on a specific
language, P4, as a case study.

\subsection{The P4 Programming Language}\label{sec:intro-p4}

P4 is a statically-typed, high-level domain-specific language (DSL) for
programming packet processors.
P4 can program a wide range of devices, from network switches to DPUs such as
NVIDIA BlueField~\cite{bluefield},
%
%
%
and can also be used as a specification language, as Google used to define the
behavior of its in-house fixed-function switches~\cite{switchv}.

P4 is evolving without a normative specification.
The P4 community maintains an official specification, a 190-page document in
informal prose~\cite{p4spec}.
The specification is official, but not normative; it is informal, incomplete,
and ambiguous.
%
The following excerpt illustrates the informality:
\begin{quote}
The conditional statement uses standard syntax and semantics familiar from many
programming languages. However, the condition expression in P4 is required to
be a Boolean (and not an integer).
\end{quote}
%
Since its initial release in 2017, P4 has incorporated
constant-valued expressions, overloading, generics, type inference, and more;
the current version reflects these accumulated features.

Evolution without a solid foundation has led to divergences across
representations.
Alongside the specification, the p4c reference compiler~\cite{p4crepo}
constitutes the P4 ecosystem.
While intending to conform to the specification, it has also grown into a giant
codebase, spanning 300K lines of C++ code.
Divergences are frequently and silently introduced.
For example, specification version 1.2.2, released in May 2021, introduced
generic types.
However, p4c failed to account for corner cases when generics interacted with
type inference, and the bug was only patched in March
2025~\cite{p4cgenericbug}.

Formalizations of P4 have also been studied~\cite{petr4, hol4p4, p4light}.
However, these do not capture the full language and remain separate from the
standardization process, effectively frozen near version 1.2.2.
Considering the role of P4 in industry and even as a specification language,
such divergence underscores the need for mechanization and its integration.

\subsection{Our Contributions}\label{sec:intro-contributions}

\begin{figure}[t]
  \centering
  \begin{tikzpicture}[
  node distance=1cm and 2cm,
  box/.style={
    draw, thick, align=left,
    font=\sffamily\footnotesize, fill=white,
    inner sep=3pt, minimum width=3.5cm
  },
  title/.style={font=\sffamily\bfseries\small},
  arrow/.style={-{Stealth[scale=1.0]}, thick}
]

\node[title] (p4_label) {P4-SpecTec};

\node[box, fill=blue!3, text width=8.2cm, minimum height=2cm] (p4_static) {
  \vspace{-0.1cm}
  \lstinputlisting[
    style=spectecstyle,
    language=spectec,
    basicstyle=\ttfamily\footnotesize,
    backgroundcolor=\color{blue!3},
    aboveskip=0pt,
    belowskip=0pt
  ]{resources/intro/p4-if-static.watsup}
  \vspace{-0.2cm}
};
\node[below=0.1cm of p4_static, align=center, font=\sffamily\small] (p4_static_caption) {
  P4 \pfourinl|if| statement typing rule
};

\node[box, fill=blue!3, text width=3.8cm, 
      right=0.1cm of p4_static.north east, anchor=north west, align=center] (p4_dynamic) {
  \vspace{-0.1cm}
  \lstinputlisting[
    style=spectecstyle,
    language=spectec,
    basicstyle=\ttfamily\footnotesize,
    backgroundcolor=\color{blue!3},
    aboveskip=0pt,
    belowskip=0pt
  ]{resources/intro/p4-if-dynamic.watsup}
  \vspace{-0.3cm}
};
\node[below=0.1cm of p4_dynamic, align=center, font=\sffamily\small] (p4_dynamic_caption) {
  P4 \pfourinl|if| statement\\dynamic semantics
};

\node[draw, dashed, gray, thick, rounded corners, 
      fit=(p4_static) (p4_static_caption) (p4_dynamic), inner sep=3pt] (envelope) {};

\node[box] (p4_spec) [below=0.4cm of envelope] {
  \textbf{P4 Spec Document (Prose)}
};
\node[box] (p4_typecheck) [left=0.15cm of p4_spec] {
  \textbf{P4 Type Checker}
};
\node[box] (p4_interp) [right=0.15cm of p4_spec] {
  \textbf{P4 Interpreter}
};

\draw[arrow] (p4_static_caption) -- node[right=7pt, font=\small, xshift=3pt, yshift=-1pt] {compile} (p4_spec);
\draw[arrow] (p4_dynamic_caption) -- node[right, font=\small, yshift=-16pt, xshift=-15pt] {compile} ([xshift=30pt]p4_spec.north);
\draw[arrow] (p4_static_caption) -- node[left, font=\small, xshift=-7pt, yshift=-1pt] {execute} (p4_typecheck);
\draw[arrow] (p4_dynamic_caption) -- node[right, font=\small] {execute} (p4_interp);

\end{tikzpicture}
  \vspace{-0.5em}
  \caption{
    (Top) Mechanization of the P4 \pfourinl|if| statement.
    (Bottom) P4-SpecTec backends.}
  \label{fig:p4-spectec}
\end{figure}
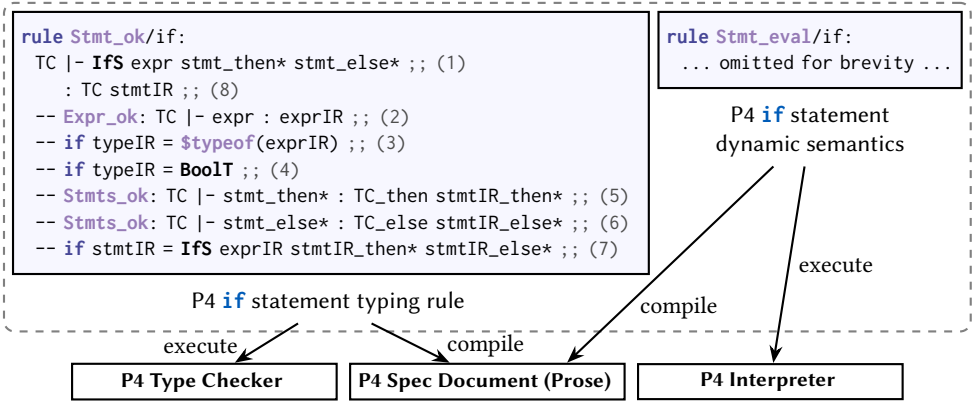

To this end, we design, implement, and integrate P4-SpecTec, a mechanization
framework for P4.
As its name suggests, P4-SpecTec builds upon Wasm-SpecTec, reusing its
meta-language syntax while \emph{redesigning the backends} to accommodate P4's
distinct features.
As shown in Fig.~\ref{fig:p4-spectec}, P4-SpecTec operates upstream, with P4
type checker, interpreter, and specification document as backends.

Revisiting the aforementioned question in Section~\ref{sec:intro-mech}, the
question can be decomposed into three parts.
To address the challenge, it is required to:
\begin{itemize}
  \item \textbf{Specify} a complete and unambiguous normative specification
    when it is absent.
  \item \textbf{Mechanize} it in a meta-language that is both executable
    and comprehensible.
  \item \textbf{Integrate} it into the language ecosystem as the
    official model for language design.
\end{itemize}
We demonstrate the concrete challenges and our solutions for each step in the
context of P4.

\vspace{2pt}
\textbf{Specification}.
Unlike JavaScript and Wasm, P4 lacks a rigorous specification.
Its semantics is (inconsistently) defined across the informal specification,
p4c compiler passes, and partial formalizations.
We reconcile these into a mechanized specification of P4.
The specification is complete with respect to version 1.2.5, excluding
concurrency, undefined values, implementation-specific features, and unresolved
semantics.
In the process, we discovered and reported \nBug{} bugs.

\vspace{2pt}
\textbf{Mechanization}.
The main technical challenge lies here, in determining the shape of the
meta-language and its semantics.
Requirements come from both the P4-SpecTec backends and P4.
To support type checker, interpreter, and document backends, the meta-language
should be readily executable and translatable into a comprehensible form, such
as prose algorithms as in JavaScript.
Simultaneously, the meta-language should accommodate the unique characteristics
of P4.

Among many distinct features of P4, its type system is the most challenging
aspect for mechanization.
Because the P4 type system computes information necessary for dynamic semantics,
e.g., cast insertion and type inference,
the mechanized type system should be executable.
However, to our knowledge, none of the aforementioned frameworks support
executability of the type system.
ESMeta is inapplicable as it targets dynamically-typed JavaScript,
and Wasm-SpecTec is inapplicable as well.
%
%
Recall in Fig.~\ref{fig:esmeta-and-wasm-spectec} that the declarative
meta-language was animated into an algorithmic form for execution.
Animation is non-trivial in general, so Wasm-SpecTec further assumes the
determinism and structural regularity of the dynamic subset of Wasm
mechanization, e.g., $\hookrightarrow$ patterns in reduction rules.
Indeed, as shown in dotted lines in the figure, Wasm-SpecTec cannot animate the
Wasm type system mechanization, due to its highly declarative and irregular
nature.

In P4-SpecTec, we choose \emph{algorithmic inference rules} as the main
instrument for mechanization.
%
Inference rules are widely understood by programming language researchers and
practitioners, useful for case analysis, and can be applied to the type system
and operational semantics alike.
We refine inference rules into algorithmic inference rules, a computational
subset that can be directly executed without the need for animation.
Further, we define the systematic steps for compiling these into prose
algorithms, for comprehension of P4 programmers.

The SpecTec meta-language is well-suited for this purpose, as its syntax
resembles textbook-style formalization.
Yet, its animation-reliant interpreter backend and non-existent type checker
backend are ill-suited for P4.
Thus, we adapt the SpecTec meta-language for specifying algorithmic inference
rules, and redesign the backends to systematically execute them and compile
them into prose.

For example, Fig.~\ref{fig:p4-spectec} shows the mechanization of the P4
\pfourinl|if| statement typing rule using algorithmic inference rules.
It resembles the textbook-style inference rule, but with the
premises with leading \mbox{\specplain{-}\specplain{-}} coming \emph{below} the
conclusion.
Here, \specplain{TC} on the left of the turnstile (\specplain{|-}) denotes the
typing context.
It reads in the order specified by the comments from 1 to 8: (1) to type an
\pfourinl|if| statement under \specplain{TC}, (2) type check \specplain{expr}
to get its typed version \specplain{exprIR}, (3-4) check that it has boolean
type, (5-6) type check both branches, (7) assemble the typed constructs into
\specplain{stmtIR}, and (8) finally result in \specplain{TC} and the typed
statement \specplain{stmtIR}.
%
%
Aligning with this interpretation, we define the constraints that
constitute algorithmic inference rules, and the systematic steps for executing
them.

Using algorithmic inference rules, we mechanize the latest version of P4,
including the type system and dynamic semantics.
Executing the mechanization yields a P4 type checker and interpreter, which are
tested against the p4c test suite, validating the completeness and correctness
of our mechanization.
The mechanization is compiled into a prose specification, to be used as the
future official specification document by the P4 community.

\vspace{2pt}
\textbf{Integration}.
%
Finally, bringing research into the real world requires more than just technology.
It amounts to a socio-technical process of understanding the governance,
building trust within the community, and demonstrating practical value of
mechanization.
Through continued engagement, the P4 Language Design Working Group has
conditionally adopted P4-SpecTec as the official toolchain for authoring the P4
specification.
Future P4 specification updates will be made at the meta-language level,
validated against the tests from the p4c test suite, and compiled for release
as a prose specification.
The adoption is conditioned on providing a documentation of the framework,
integrating P4-SpecTec into the CI system of the P4 ecosystem, and having a
transitional period for the P4 community to adapt to the new specification.

\vspace{2pt}
Our work, P4-SpecTec, is the first end-to-end case study of real-world
adoption of mechanization for a language lacking a normative specification.
Our contributions in this work are as follows:
\begin{itemize}
  \item We present the first executable mechanization of P4 that covers the
    syntax, type system, and dynamic semantics, up to the latest version,
    1.2.5.
  \item We define \emph{algorithmic inference rules} as a mechanization instrument.
  We introduce the systematic steps for executing algorithmic inference
    rules into a P4 type checker and interpreter, and compiling them into prose
    specification (\cref{sec:algorithmic}, \cref{sec:structured}, \cref{sec:prose}).
  \item We share the lessons learned from integrating P4-SpecTec into the P4
    ecosystem (\cref{sec:integration}).
  \item Using P4-SpecTec, we identify \nBug{} bugs across P4 representations
    (\cref{sec:eval}).
\end{itemize}
%
%


\section{Background}\label{sec:background}

In this section, we provide background on the P4 language
(\cref{sec:background:p4}) and its ecosystem (\cref{sec:background:eco}).

\subsection{The P4 Language}\label{sec:background:p4}

\begin{figure}[t]
  \centering
  \begin{minipage}[t]{0.48\textwidth}
    \vspace{0pt}
    \centering
    \resizebox{\textwidth}{!}{
      \begin{tikzpicture}[
  block/.style={
    rectangle, 
    draw=black, 
    thick, 
    fill=blue!5, 
    minimum height=0.5cm, 
    align=center,
    font=\sffamily\bfseries
  },
  groupbox/.style={
    draw,
    dashed,
    inner sep=0.1cm,
    rounded corners=2pt
  },
  arrow/.style={
    -{Stealth[scale=1.2]}, 
    thick
  },
  line/.style={
    thick
  }
]

\node[block] (parser) {Parser};
\node[block, above=0.7cm of parser.north west, anchor=south west, fill=red!5] (packetin) {Input Engine};
\node[block, right=0.5cm of parser] (processor) {Match-action(s)};
\node[block, right=0.5cm of processor] (deparser) {Deparser};
\node[block, above=0.7cm of deparser.north east, anchor=south east, fill=red!5] (packetout) {Output Engine};

\node[groupbox, fit=(parser) (processor) (deparser) (packetin) (packetout)] (container) {};

\node[block, draw, dotted, fill=gray!10, above=1.7cm of processor] (cp) {Control Plane};

\node[left=0.5cm of parser, align=center] (in) {Input\\Packet};
\node[right=0.5cm of deparser, align=center] (out) {Output\\Packet};
\node[below=0.1cm of out] (drop) {Drop};

\draw[arrow] (packetin) -- (parser);
\draw[arrow] (parser) -- (packetin);
\draw[arrow] (packetout) -- (deparser);
\draw[arrow] (deparser) -- (packetout);

\draw[arrow] (in) -- (parser);
\draw[arrow] (parser) -- (processor);
\draw[arrow] (processor) -- (deparser);
\draw[arrow] (deparser) -- (out);
\draw[arrow] (processor.south) |- (drop);
\draw[arrow, dashed] (cp) -- node[right, yshift=-14pt] {populate} (processor);

\end{tikzpicture}
    }
    \vspace{-2.3em}
    \caption{VSS architecture diagram.}
    \label{fig:background:p4:arch}
    \vspace{0.5em}
\begin{lstlisting}[
      style=p4style,
      language=p4,
      basicstyle=\ttfamily\scriptsize,
      xleftmargin=1.5em,
    ]
extern packet_in {
  void extract<T>(out T hdr);
}
extern packet_out {
  void emit<T>(in T hdr);
}
struct Meta {
  bit<8> inport; bit<8> outport; bool drop;
}
parser Parse<H>(packet_in p, out H hdr,
                inout Meta meta);
control Control<H>(inout H hdr,
                   inout Meta meta);
control Deparse<H>(packet_out p, in H hdr);
package VSS<H>(Parse<H> parse,
               Control<H> ctrl,
               Deparse<H> deparse);
\end{lstlisting}
    \vspace{-1em}
    \caption{\texttt{VSS} architecture header file in P4 (\progname{vss.p4}).}
    \label{fig:background:p4:vss}
  \end{minipage}
\hspace*{1.1em}
  \begin{minipage}[t]{0.48\textwidth}
    \vspace{-0.5em}
\begin{lstlisting}[
      style=p4style,
      language=p4,
      basicstyle=\ttfamily\scriptsize,
      xleftmargin=1.5em,
    ]
#include <vss.p4>
header Hdr { bit<8> src; bit<8> dst; }

parser Prs(packet_in p, out Hdr hdr,
           inout Meta meta) {
  state start {
    p.extract(hdr);
    transition accept;
  }
}
control Ctrl(inout Hdr hdr,
             inout Meta meta) {
  action allow() { meta.outport = 1; }
  action drop() { meta.drop = true; }
  table access_control {
    key = { hdr.src: exact; }
    actions = { allow; drop; }
  }
  apply { access_control.apply(); }
}
control Dep(packet_out p, in Hdr hdr) {
  apply {
    if (hdr.src != hdr.dst) p.emit(hdr);
    return;
  }
}

VSS(Prs(), Ctrl(), Dep()) main;
\end{lstlisting}
    \vspace{-1em}
    \caption{A program for \texttt{VSS} architecture (\progname{switch.p4}).}
    \label{fig:background:p4:switch}
  \end{minipage}
\vspace*{-1em}
\end{figure}

\paragraph{Targets}
P4 is a DSL designed to program diverse packet processing targets, including
switches, DPUs, and even the Linux kernel's eBPF datapath~\cite{p4cebpf}.
P4 provides programmability over their common abstraction: custom header
definitions, programmable header parsing, and packet processing based on
match-action tables.
%
Consider a Very Simple Switch (VSS), shown in Fig.~\ref{fig:background:p4:arch}.
VSS has a three-stage pipeline architecture, consisting of a packet parser,
match-action table(s), and a deparser.
Packets enter the pipeline, where they are either emitted after deparsing or
dropped during match-action.
The three stages are programmable via P4.
Besides the programmable components, each target also has its specific
fixed-function components.
For instance, the packet input/output engines in
Fig.~\ref{fig:background:p4:arch} are fixed-function components responsible for
low-level packet I/O and are not programmable by P4.
The target-specific details are abstracted as \emph{architectures} in P4.
An architecture is abstractly defined as a header file, as in \progname{vss.p4}
in Fig.~\ref{fig:background:p4:vss}.
It declares the interfaces of the \pfourinl|extern|s for packet I/O
(\pfourinl|packet_in|, \pfourinl|packet_out|), and the type signatures of the
programmable blocks to be implemented (\pfourinl|Parse|, \pfourinl|Control|,
\pfourinl|Deparse|).
P4 programs for the architecture \pfourinl|#include| the header file.
We now explain \progname{switch.p4} in Fig.~\ref{fig:background:p4:switch}, a
simple access control program implementing \progname{vss.p4}.

\vspace*{-.5em}
\paragraph{Packet processing abstraction}
%
%
At the beginning of \progname{switch.p4}, we define a custom header
\pfourinl|Hdr|, with fields \pfourinl|src| and \pfourinl|dst| of 8-bit
unsigned integer type.
\pfourinl|Prs| implements \pfourinl|Parse| as a finite-state machine, starting
from the \pfourinl|start| state.
Using the extern method call \pfourinl|p.extract(hdr)| in line 7, the parser
extracts 16 bits from the input packet and populates the fields of
\pfourinl|hdr|.
\pfourinl|Ctrl| implements \pfourinl|Control|, containing a match-action table
\pfourinl|access_control|.
It matches on \pfourinl|hdr.src|, and performs the \pfourinl|allow| or
\pfourinl|drop| action, setting the output port to \pfourinl|1|, or dropping
the packet, respectively.
In \pfourinl|drop|, \pfourinl|meta.drop| is set to \pfourinl|true|, which
informs the target to drop the packet after \pfourinl|Ctrl| execution.
%
\pfourinl|Dep| implements \pfourinl|Deparse|.
It
emits nothing if \pfourinl|hdr.src| and \pfourinl|hdr.dst| are equal, and
otherwise emits the header using \pfourinl|p.emit(hdr)|.
Emitting nothing is functionally equivalent to a packet drop, but we use an
\pfourinl|if| here for illustration purposes.
The three programmable blocks are packaged into \pfourinl|main| at the end.

\vspace*{-.5em}
\paragraph{P4 type system}
A P4 program is type checked before execution.
In addition to the ``usual'' type checking of an imperative language with
user-defined types, the type checker also makes explicit the information
required for execution.
Note that while the signature of \pfourinl|VSS| in \progname{vss.p4} is
generic, \pfourinl|main| in line 28 of \progname{switch.p4} omits the type
argument for parameter \pfourinl|H|.
The type checker is responsible for inferring the omission as
\pfourinl|VSS<Hdr>(...)|.
At runtime, the target is informed of the concrete header type to be parsed,
processed, and deparsed by consulting the inferred type arguments.

\vspace*{-.5em}
\paragraph{Instantiation}
\pfourinl|access_control| is a table that matches on the specified keys and
performs one of the specified actions.
Its concrete table entries are not specified in \progname{switch.p4}.
They are populated dynamically by the control plane, a network controller
separate from the device running P4.
Hence, \pfourinl|table|s are stateful, where updates from the control plane
take effect across multiple packets.
Externs are stateful in general as well.
Consider \pfourinl|p.extract(hdr)| in line 7.
Calling the method once differs from calling it twice; each call advances the
packet cursor.
P4 defines a static instantiation phase between type checking and execution,
during which all stateful resources are allocated.
Starting from the \pfourinl|main| declaration in line 28, all stateful
resources are recursively instantiated.

\vspace*{-.5em}
\paragraph{Open-world semantics}
Notice that \progname{switch.p4} does \emph{not} have a clear entry point,
since neither \pfourinl|Prs| nor \pfourinl|Ctrl| is explicitly invoked.
Also, the behavior of the \pfourinl|packet_in| and \pfourinl|packet_out|
externs is not defined in \progname{switch.p4}, but instead provided as
fixed-function components of the target.
Hence, a P4 program alone does not fully determine the semantics of the entire
system in Fig.~\ref{fig:background:p4:arch}.
The target is responsible for providing the entry point, orchestrating the
execution of the programmable components, and supplying the extern
implementations.
Therefore, P4-SpecTec must allow interaction between the mechanized P4
semantics and the external semantics of the target.

In summary, P4-SpecTec must support (i) execution of the P4 type system, (ii)
recursive instantiation of stateful resources, and (iii) interaction with
external semantics.

\subsection{The P4 Language Ecosystem}\label{sec:background:eco}

\begin{table}[t]
  \centering
  \footnotesize
  \setlength{\aboverulesep}{0pt}
  \setlength{\belowrulesep}{0pt}
  \renewcommand{\arraystretch}{0.9}
  \caption{Comparison of P4 formalization efforts and P4-SpecTec}
  \label{tab:background:eco:compare}
  \vspace{-1em}
  \begin{tabular}{lccc|c}
  \toprule
   & \textbf{Petr4} & \textbf{P4Light} & \textbf{HOL4P4} & \textbf{P4-SpecTec} \\ \midrule
  \textbf{Mechanization} & LaTeX & \multirow{2}{*}{Rocq} & Ott & \multirow{2}{*}{SpecTec} \\
  \textbf{Implementation} & OCaml & & HOL4{\textsuperscript{\dag}} & \\ \midrule
  \textbf{Type system} & Y & N & Y & Y \\
  \textbf{Static instantiation} & N & Y & N{\textsuperscript{*}} & Y \\
  \textbf{Dynamic semantics} & Big-step & Big-step & Small-step & Big-step \\
  \textbf{Concurrency} & N & N & Y & N \\ \bottomrule
  \end{tabular}
\\
{\scriptsize \textsuperscript{\dag} Derived from the specification in Ott, only for the dynamic semantics.
 \textsuperscript{*} Does not support recursive instantiation.}
\vspace*{-1em}
\end{table}

The P4 language ecosystem is primarily shaped by the official specification and
the p4c reference compiler.
The specification~\cite{p4spec} is an open-source AsciiDoc document describing
P4 in informal prose.
The P4 Language Design Working Group (LDWG)~\cite{p4ldwg} maintains the specification.
LDWG is open to the public, where discussions and decisions take place on
GitHub issues and pull requests, and in monthly virtual meetings.
First published in May 2017 as version 1.0.0, the specification has been
updated roughly annually, with the latest version, 1.2.5, published in October
2024.
%

p4c~\cite{p4crepo} is also open-source.
Alongside its C++ codebase, the repository contains a test suite
with regression tests for reported bugs.
The compiler has evolved with the specification, intended to adhere to it.
The compiler may deliberately deviate from the specification for practical
reasons such as optimization and experimental features.
%
However, due to the language's growing complexity, unintended divergences have
silently accumulated.
Moreover, because p4c adopts a nano-pass compiler architecture~\cite{p4cpaper},
it often fails to account for subtle interactions between different features.
%
We report such divergences between p4c and the
specification identified using P4-SpecTec (\cref{sec:eval}).

The lack of a clear specification has motivated P4 formalization efforts,
including Petr4~\cite{petr4}, P4Light~\cite{p4light}, and HOL4P4~\cite{hol4p4},
%
as summarized in Table~\ref{tab:background:eco:compare}.
Petr4 presented the first P4 formalization, covering its type system and
dynamic semantics in big-step.
P4Light is a Rocq formalization focused on instantiation and verifying the
properties of stateful resources in P4.
%
%
HOL4P4 formalizes P4 in small-step using the Ott framework~\cite{ott}, focusing
on the concurrent semantics of P4.
Despite these efforts, they define only core fragments of P4 and are not up to
date with the latest specification.

P4-SpecTec builds on these artifacts to provide better coverage of P4 features,
up to version 1.2.5.
%
%
As Table~\ref{tab:background:eco:compare} shows,
P4-SpecTec covers the type system and instantiation, which others support only
partly.
Because it uses big-step semantics, it does not support concurrency.
Nevertheless, we argue that this is not a major limitation, since discussions
on concurrency remain at an early stage within the LDWG, where even the notion
of an atomic step is still an open question.
We also present an empirical comparison of these formalizations and P4-SpecTec
using the p4c test suite (\cref{sec:eval}).

\section{Overview of P4-SpecTec}\label{sec:overview}

\subsection{The SpecTec Meta-language}\label{sec:overview-metalang}

\begin{figure}[t]
  \centering
  \begin{minipage}[t]{0.48\textwidth}
    \vspace{0pt}
\begin{lstlisting}[
      style=spectecstyle,
      language=spectec,
      basicstyle=\ttfamily\footnotesize,
      backgroundcolor=\color{blue!3},
    ]
@\emptyline@
 @\speckey{syn}@ id = @\speckey{text}@
 @\speckey{syn}@ type = | @\specterm{BoolT}@ | @\specterm{BitT}@ | @\specterm{VarT}@ id
 @\speckey{syn}@ expr =
  | @\specterm{VarE}@ id | @\specterm{BoolE}@ @\speckey{bool}@ | @\specterm{BitE}@ @\speckey{nat}@
  | @\specterm{CmpE}@ expr expr | @\specterm{FieldE}@ expr id
 @\speckey{syn}@ stmt =
  | @\specterm{IfS}@ expr stmt* stmt* | @\specterm{RetS}@ expr?
  | @\specterm{CallS}@ id id type* expr*
 @\speckey{syn}@ method, param = ... omitted ...
 @\speckey{syn}@ decl =
  | @\specterm{ExternD}@ id method* | @\specterm{HeaderD}@ id (id, type)*
  | @\specterm{ControlD}@ id param* stmt*
 @\speckey{syn}@ pgm = decl*
\end{lstlisting}
    \vspace{-1em}
    \caption{P4 abstract syntax in P4-SpecTec (simplified).}
    \label{fig:p4-syntax}
  \end{minipage}
  \hfill
  \begin{minipage}[t]{0.49\textwidth}
    \vspace{0pt}
\begin{lstlisting}[
      style=spectecstyle,
      language=spectec,
      basicstyle=\ttfamily\footnotesize,
      backgroundcolor=\color{blue!3},
    ]
@\emptyline@
 @\speckey{syn}@ methodIR = ... omitted ...
 @\speckey{syn}@ typeIR =
  | @\specterm{BoolT}@ | @\specterm{BitT}@ | @\specterm{ExternT}@ id methodIR*
  | @\specterm{HeaderT}@ id (id, typeIR)*
 @\speckey{syn}@ exprIR =
  | @\specterm{VarE}@ id typeIR | @\specterm{BoolE}@ bool typeIR
  | @\specterm{BitE}@ nat typeIR | @\specterm{CmpE}@ exprIR exprIR typeIR
  | @\specterm{FieldE}@ exprIR id typeIR
 @\speckey{syn}@ stmtIR =
  | @\specterm{IfS}@ exprIR stmtIR* stmtIR* | @\specterm{RetS}@ exprIR?
  | @\specterm{CallS}@ id id typeIR* exprIR*
 @\speckey{syn}@ declIR = ... omitted ...
 @\speckey{syn}@ pgmIR = declIR*
\end{lstlisting}
    \vspace{-1em}
    \caption{P4IR syntax in P4-SpecTec (simplified).}
    \label{fig:p4ir-syntax}
  \end{minipage}
  \\[1em]
  \begin{minipage}[t]{0.43\textwidth}
    \vspace{0pt}
\begin{lstlisting}[
      style=spectecstyle,
      language=spectec,
      basicstyle=\ttfamily\footnotesize,
      backgroundcolor=\color{blue!3},
    ]
@\emptyline@
 @\speckey{syn}@ TC = ... typing context ...
 @\speckey{rel}@ @\specrel{Expr\_ok}@: TC |- expr : exprIR
    @\speckey{hint}@(input %0 %1)
 @\speckey{rel}@ @\specrel{Stmt\_ok}@: TC |- stmt : TC stmtIR
    @\speckey{hint}@(input %0 %1)
 @\speckey{rel}@ @\specrel{Pgm\_ok}@: |- pgm : pgmIR @\speckey{hint}@(input %0)
@\emptyline@
 @\speckey{ext syn}@ state
 @\speckey{syn}@ obj = @\specterm{ExternO}@ state | ...
 @\speckey{syn}@ STO = ... store holding objs ...
 @\speckey{rel}@ @\specrel{Pgm\_inst}@: |- pgmIR : STO
    @\speckey{hint}@(input %0)
\end{lstlisting}
    \vspace{-1em}
    \caption{P4 typing and instantiation (simplified).}
    \label{fig:p4-static-rel}
  \end{minipage}
  \hfill
  \begin{minipage}[t]{0.545\textwidth}
    \vspace{0pt}
\begin{lstlisting}[
      style=spectecstyle,
      language=spectec,
      basicstyle=\ttfamily\footnotesize,
      backgroundcolor=\color{blue!3},
    ]
@\emptyline@
 @\speckey{syn}@ EC = ... eval context ...
 @\speckey{syn}@ val = @\specterm{BoolV}@ @\speckey{bool}@ | ...
 @\speckey{syn}@ sign = @\specterm{CONT}@ | @\specterm{RET}@ val?
@\emptyline@
 @\speckey{rel}@ @\specrel{Expr\_eval}@: EC |- exprIR : val
    @\speckey{hint}@(input %0 %1)
 @\speckey{rel}@ @\specrel{Stmt\_eval}@: STO EC |- stmtIR : STO EC sign
    @\speckey{hint}@(input %0 %1 %2)
 @\speckey{rel}@ @\specrel{Call\_p4}@: STO EC |- id val* : STO EC
    @\speckey{hint}@(input %0 %1 %2 %3)
 @\speckey{ext rel}@ @\specrel{Call\_ext}@: state |- id id typeIR* val* : state
    @\speckey{hint}@(input %0 %1 %2 %3 %4)
\end{lstlisting}
    \vspace{-1em}
    \caption{P4 evaluation (simplified).}
    \label{fig:p4-eval-rel}
  \end{minipage}
\vspace*{-.5em}
\end{figure}

%
The P4-SpecTec meta-language syntax is largely the same as in Wasm-SpecTec.
%
However, the semantics of the meta-language is redesigned in
P4-SpecTec to accommodate requirements like executable type system and open-world semantics.
Three meta-language constructs allow for defining a language: syntax, relation,
and function.
%
%
We explain them using \pfourinl|control Dep| in
Fig.~\ref{fig:background:p4:switch} as an example.
Note that there are two languages in play: the P4 language being specified, and
the P4-SpecTec meta-language.
To disambiguate, we prefix \emph{meta-} when referring to the meta-language.

\vspace*{-.5em}
\paragraph{P4 Syntax}
In Fig.~\ref{fig:p4-syntax}, we show the simplified abstract syntax of P4
defined in P4-SpecTec, capturing \pfourinl|control Dep|.
\speckeyinl{syn} introduces a user-defined meta-type, where \speckeyinl{text}
and \speckeyinl{nat} are primitive meta-types for strings and natural numbers,
respectively.
The defined meta-type can be an alias for an existing meta-type
\mbox{(\speckeyinl{syn}\specplain{ id})} or a variant type
\mbox{(\speckeyinl{syn}\specplain{ type})}.
Starting from the root non-terminal \specplain{pgm}, P4 is defined along three
main syntactic categories: \specplain{decl}, \specplain{stmt}, and
\specplain{expr}.
Variants are defined with a leading constructor, where a \specplain{stmt} is
either a conditional (\specterminl{IfS}), a return
(\specterminl{RetS}), or a method call (\specterminl{CallS}).
The fields of \specterminl{CallS} represent the \specplain{id} of the object,
the \specplain{id} of the method, type arguments \specplain{type*} and
arguments \specplain{expr*}.
\specplain{*} and \specplain{?} represent iterators as in EBNF.
\specplain{expr*} in \specterminl{CallS} represents a (possibly empty) list of
\specplain{expr}s, and \specplain{expr?} in \specterminl{RetS} represents an
optional \specplain{expr} for the return value.
However, in the actual mechanization, we specify the concrete syntax of P4:
%
\begin{lstlisting}[
  style=spectecstyle,
  language=spectec,
  basicstyle=\ttfamily\footnotesize,
  backgroundcolor=\color{blue!3},
]
 @\speckey{syn}@ conditionalStatement =
   | @\specterm{IF}@ @\specterm{(}@ expression @\specterm{)}@ statement | @\specterm{IF}@ @\specterm{(}@ expression @\specterm{)}@ statement @\specterm{ELSE}@ statement
\end{lstlisting}
%
Unlike in the abstract syntax, terminal symbols and non-terminals are
interleaved.
%

\vspace*{-.5em}
\paragraph{P4 Intermediate Representation (IR)}
Because the P4 type system does more than simple type checking, such as making
implicit casts explicit and type inference, we define P4IR that extends the
surface P4 syntax.
Hence, the mechanized P4 type system takes a P4 program as input, and produces
a typed P4IR program if the program is well-typed.
For instance, line 23 in \progname{switch.p4} omits the type argument of the
\pfourinl|emit| method call, which is made explicit as
\pfourinl|p.emit<Hdr>(hdr);| in P4IR.
Later instantiation and dynamic semantics are defined on the P4IR.
Fig.~\ref{fig:p4ir-syntax} shows the P4IR syntax, close to the surface syntax
in Fig.~\ref{fig:p4-syntax}.
%
%
For \specplain{type}, \specterminl{VarT} is resolved to the user-defined P4
types, \specterminl{ExternT} and \specterminl{HeaderT} in \specplain{typeIR}.
%
%
In \specplain{exprIR}, variants are annotated with their types, such as
\mbox{\specplain{(}\specterminl{NameE}\specplain{ id typeIR)}}.

\vspace*{-.5em}
\paragraph{Relations}
Inference rules are the main way of defining semantics.
A \speckeyinl{rel} declares the meta-type signature of a set of inference
rules.
In Fig.~\ref{fig:p4-static-rel}, we show the relations for typing and
instantiation.
\specrelinl{Stmt\_ok} is a relation for typing statements, defined on a typing
context \specplain{TC}, the \specplain{stmt} to be typed, and the updated
\specplain{TC} and \specplain{stmtIR} if the statement is well-typed.
The turnstile and colon (\specplain{|-}, \specplain{:}) are notational symbols
to resemble the textbook-style presentation, without a specific semantics.
In P4-SpecTec, users must also provide \mbox{\speckeyinl{hint}\specplain{(input
...)}} for each \speckeyinl{rel}, indicating which parts are input to the
relation, and which are output.
For \specrelinl{Stmt\_ok}, the hint specifies that the zero-th (\specplain{TC})
and the first (\specplain{stmt}) are the inputs, and the rest are outputs.
%
Mathematically, a relation is a set of tuples, but the hint is required to
execute inference rules algorithmically.
\speckeyinl{rel}s in Fig.~\ref{fig:p4-static-rel} and
Fig.~\ref{fig:p4-eval-rel} reveal how P4 programs are typed, instantiated, and
evaluated.
%
\specrelinl{Pgm\_ok} takes a P4 program \specplain{pgm} and produces a
typed \specplain{pgmIR}.
Then, \specrelinl{Pgm\_inst} takes \specplain{pgmIR} and produces a store
(\specplain{STO}) containing the instantiated program.
Within \specplain{STO}, there are objects (\specplain{obj}) such as
\specterminl{ExternO} for instantiated extern objects.
\specrelinl{Call\_p4} calls the programmable blocks in \specplain{STO}, which
takes the block to be called (\specplain{id}) and the inputs
(\specplain{val*}).
From \specrelinl{Call\_p4}, through a series of rule applications,
\specrelinl{Stmt\_eval} evaluates a P4 statement in the block and produces
\specplain{sign}, an indicator for return (\specterminl{RET}) or continue
(\specterminl{CONT}).

\vspace*{-.5em}
\paragraph{Inference Rules and Meta-functions}
\speckeyinl{rule}s specify each semantic case of a \speckeyinl{rel}.
Fig.~\ref{fig:p4-spectec} shows the typing rule for \pfourinl|if| statements.
Its presentation follows the standard notational conventions of inference
rules, but with the premises appearing \emph{below} the conclusion.
The rule is read as indexed in the comments, starting from the input
(\specplain{TC} and \specterminl{IfS}), through the premises in order, and
finally reaching the conclusion.
%
Notice how this order of interpretation intuitively suggests how the rule
should be executed, which is generally not the case for general declarative
inference rules.
Building on this intuition, we define the constraints on inference rules that
make them algorithmic, and how they are executed in
Section~\ref{sec:algorithmic-constraints}.
Besides relations, auxiliary meta-functions may also be defined as:
\begin{lstlisting}[
  style=spectecstyle,
  language=spectec,
  basicstyle=\ttfamily\footnotesize,
  backgroundcolor=\color{blue!3},
]
 @\speckey{dec}@ @\specfunc{\$typeof}@(exprIR) : typeIR
 @\speckey{def}@ @\specfunc{\$typeof}@(@\specterm{NameE}@ id typeIR) = typeIR        @\speckey{def}@ @\specfunc{\$typeof}@(...) = ... omitted ...
\end{lstlisting}
\specfuncinl{\$typeof} is declared with the \speckeyinl{dec} keyword and
implemented case by case with the \speckeyinl{def} keyword.

\vspace*{-.5em}
\paragraph{Extern Interface}
Syntax, relations, and meta-functions can be defined with a leading
\speckeyinl{ext} keyword, as in
\mbox{\speckeyinl{ext syn}\specplain{ state}} in Fig.~\ref{fig:p4-static-rel}
and \mbox{\speckeyinl{ext rel}\specplain{ }\specrelinl{Call\_ext}} in
Fig.~\ref{fig:p4-eval-rel}.
\speckeyinl{ext} is unique to P4-SpecTec, which introduces external components
into the mechanization.
\specplain{state} represents an internal state of an external object
implementation, which is outside the scope of the core P4 semantics.
\specrelinl{Call\_ext} is a relation for calling an external method, also
external to the core P4 semantics.
Hence, these are only declared, but their details are not defined in the
mechanization.
We leave its details to Section~\ref{sec:algorithmic:extern}.

\vspace*{-.5em}
\paragraph{Iteration}
P4-SpecTec supports two iterators: \specplain{*} and \specplain{?}.
In Fig.~\ref{fig:p4-spectec}, the meta-expression \specplain{stmt\_then*}
represents the list of statements in the then-branch of an \pfourinl|if|
statement.
%
General meta-expressions and premises can be iterated.
Suppose we define an \pfourinl|if| statement as
\mbox{\specterminl{IfS}\specplain{ expr* stmt* stmt*}}, where the condition is
a list of \specplain{expr}s.
%
Typing those is a premise
\mbox{\specplain{(}\specrelinl{Expr\_ok}\specplain{: TC |- expr : exprIR)*}},
meaning that for each \specplain{expr} in \specplain{expr*},
\specrelinl{Expr\_ok} is applied to
derive \specplain{exprIR}.
\specplain{exprIR} are then collected into a list, \specplain{exprIR*}.
Similarly,
\mbox{\specfuncinl{\$typeof}\specplain{(exprIR)*}} maps over the list of
\specplain{exprIR}, yielding a list of \specplain{typeIR}.
Iterators may also be nested, such as \specplain{x?*} for a list of optional
\specplain{x} and \specplain{y**} for a 2D list of \specplain{y}.

\subsection{P4-SpecTec Architecture}

\begin{figure}[t]
  \centering
  \input{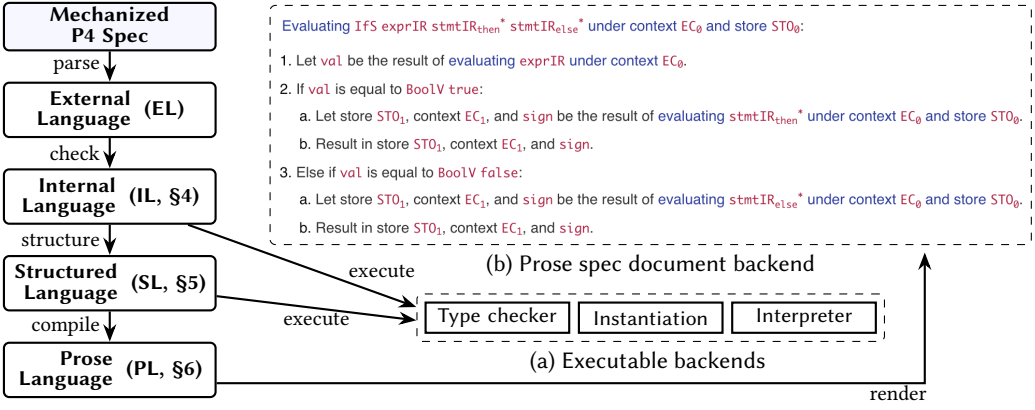}
\vspace*{-.5em}
  \caption{
    Overview of the P4-SpecTec architecture.
  }
  \label{fig:p4-spectec-arch}
\vspace*{-.5em}
\end{figure}

Fig.~\ref{fig:p4-spectec-arch} shows the overview of P4-SpecTec, which has a
multi-stage architecture, taking a mechanized specification at the top as
input.
%
It (i) type checks and validates the algorithmicity of the inference rules,
(ii) adds structured control flow to the rules, and (iii) compiles the rules
into prose algorithms.
Each stage refines the specification so that it can be executed as in (a) and
read by P4 specification readers as in (b).
First, the input specification is parsed into an \emph{EL (External Language)}
representation.
Then, EL is meta-type checked and elaborated into a more explicit \emph{IL
(Internal Language)}.
During the process, iterated meta-expressions and premises are annotated and
validated of their iterators and dimensions.
%
Assume we have meta-variables \specplain{x*} and \specplain{y**}.
In elaboration, iterated meta-expression such as
\mbox{\specplain{(}\specfuncinl{\$f}\specplain{(x,1))*}} is annotated that it
applies \mbox{\specfuncinl{\$f}\specplain{(x,1)}} for each \specplain{x} in
\specplain{x*}.
Iteration dimensions are checked for consistency, e.g.,
\mbox{\specplain{(x+y)*}} is invalid as it omits the second dimension of
\specplain{y**}.
These are mostly the same as in Wasm-SpecTec.
The subsequent stages are redesigned in P4-SpecTec.

\begin{figure}[t]
  \centering
\begin{lstlisting}[
    style=spectecstyle,
    language=spectec,
    basicstyle=\ttfamily\footnotesize,
    backgroundcolor=\color{blue!3},
  ]
@\emptyline@
 @\speckey{rule}@ @\specrel{Stmt\_eval}@/if-then:
   STO_0 EC_0 |- @\specterm{IfS}@ exprIR stmtIR_then* stmtIR_else* : STO_1 EC_1 sign
   -- @\specrel{Expr\_eval}@: EC_0 |- exprIR : @\specterm{BoolV}@ @\speckey{true}@
   -- @\specrel{Stmts\_eval}@: STO_0 EC_0 |- stmtIR_then* : STO_1 EC_1 sign
@\emptyline@
 @\speckey{rule}@ @\specrel{Stmt\_eval}@/if-else:
   STO_0 EC_0 |- @\specterm{IfS}@ exprIR stmtIR_then* stmtIR_else* : STO_1 EC_1 sign
   -- @\specrel{Expr\_eval}@: EC_0 |- exprIR : @\specterm{BoolV}@ @\speckey{false}@
   -- @\specrel{Stmts\_eval}@: STO_0 EC_0 |- stmtIR_else* : STO_1 EC_1 sign
\end{lstlisting}   
  \vspace{-1em}
  \caption{P4 \pfourinl|if| statement evaluation rules.}
  \label{fig:p4-eval-rule}
\end{figure}

\emph{Algorithmic inference rules} are the overarching concept in P4-SpecTec,
differentiating it from Wasm-SpecTec.
These are designed to be directly executable, without the need for non-trivial
translation such as animation in Wasm-SpecTec.
Furthermore, we compile the rules into a prose algorithm, comprehensible to a
broad P4 specification audience.
For instance, rules \specrelinl{Stmt\_eval}\specplain{/if-then} and
\specplain{if-else} in Fig.~\ref{fig:p4-eval-rule} are compiled into the prose
algorithm in Fig.~\ref{fig:p4-spectec-arch} (b).

To support algorithmic inference rules, the framework performs additional
validation during the EL-to-IL stage.
It validates that the inference rules in EL are indeed algorithmic.
The precise definition of algorithmic inference rules and the validation
process are described in Section~\ref{sec:algorithmic}.
Once validated, the specification in IL can be executed.
The IL interpreter executes the rules by backtracking, trying to apply each
rule until one of them applies successfully.
This effectively yields a P4 type checker, instantiation engine, and
interpreter in Fig.~\ref{fig:p4-spectec-arch} (a).

IL is further processed to \emph{SL (Structured Language)}, where control flow
structures are inserted into the specification.
An inference rule, by construction, does not contain control flow.
Each rule only specifies a single case.
This introduces redundancy, where multiple rules have similar premises but
different conclusions.
For instance, the \specrelinl{Stmt\_eval}\specplain{/if-then} and
\specplain{if-else} rules in Fig.~\ref{fig:p4-eval-rule} both evaluate
\specrelinl{Expr\_eval} for \specplain{exprIR} in the first premise, but differ
in the other premises and conclusions.
The redundancy makes execution inefficient, as the same premises are evaluated
multiple times during backtracking.
Redundancy is also not ideal for generating concise prose algorithms.
For readers, it is more intuitive to read the two rules as a whole, where the
condition is evaluated, then branching occurs based on the condition value.
%
%
Thus, SL merges the rules into a structured form, where redundancy is reduced
and branching is made explicit.
In the generated prose in Fig.~\ref{fig:p4-spectec-arch} (b), the two rules are
merged, with the first step in common and branching in steps 2 and 3.
SL can also be executed, deriving the executable backends as with IL.
Details of SL are described in Section~\ref{sec:structured}.

The final stage is \emph{PL (Prose Language)}, where we compile SL into prose
algorithms.
SL is already algorithmic and has control structures, thus the compilation is
mostly straightforward.
For readability of the prose, we allow users to annotate the input
specification with prose hints, which are inlined into the generated prose.
The prose algorithms are output in AsciiDoc, then rendered as either HTML or
PDF documents.
Details of PL are described in Section~\ref{sec:prose}.

\section{Algorithmic Inference Rules}\label{sec:algorithmic}

In this section,
%
we define the constraints that algorithmic inference rules must satisfy
(\cref{sec:algorithmic-constraints}), explain how they are executed
(\cref{sec:algorithmic:exec}), and describe how they interface with external
semantics (\cref{sec:algorithmic:extern}).

\subsection{Validation for Algorithmic Inference Rules}\label{sec:algorithmic-constraints}

Recall the intuitive interpretation of an algorithmic inference rule in
Fig.~\ref{fig:p4-spectec}.
Reading it from the input, through the ordered premises, and finally to the
output, it suggests an algorithmic reading.

\vspace*{.5em}
\begin{lstlisting}[
  style=spectecstyle,
  language=spectec,
  basicstyle=\ttfamily\footnotesize,
  backgroundcolor=\color{blue!3},
  aboveskip=0pt,
  belowskip=0pt
]
@\emptyline@
 @\speckey{rule}@ @\specrel{Type\_sub}@/trans: C |- type_a <: type_b
   -- @\specrel{Type\_sub}@: C |- type_a <: type_c
   -- @\specrel{Type\_sub}@: C |- type_c <: type_b
\end{lstlisting}
\noindent
%
However, general inference rules are declarative and do not necessarily specify
how their conclusions should be derived algorithmically.
For example, the transitivity rule for subtyping, which appears in many type
systems including the Wasm type system above, concludes that
\specplain{type\_a} is a subtype of \specplain{type\_b} if there exists some
\specplain{type\_c}.
However, the rule itself does not specify how such a \specplain{type\_c} should
be found.
Because the meta-variable is \emph{existential}, the rule is not directly
executable.

We first briefly clarify the semantics of inference rules.
Viewed as a proof system, inference rules define the cases used to construct
proof trees.
Given a desired conclusion--e.g., that an \pfourinl|if| statement is
well-typed--the semantics is defined by searching for a proof tree that derives
the conclusion, where each node corresponds to an inference rule application.
Such proof search is generally non-algorithmic, because it may require guessing
the existential choice of meta-variables (e.g., \specplain{type\_c}) and may
also be non-deterministic when multiple trees exist.

Algorithmic inference rules form a subset of inference rules where such search
is computable.
%
This is the main mechanism
to specify the P4 semantics in case-by-case, while ensuring executability and
enabling compilation to prose algorithms.
%
Algorithmic inference rules satisfy the following:
\begin{itemize}
  \item \textbf{Dataflow.} The search must \emph{not} require witnesses for
    existential meta-variables. Instead, the dataflow must be explicit,
    computing new values from already known ones.
  \item \textbf{Determinism}. Rules must be deterministic. Semantics that may
    produce multiple outputs for the same input is undesirable in P4.
\end{itemize}

\begin{figure}[t]
  \centering
\vspace*{-.5em}
  \begin{minipage}[t]{0.48\textwidth}
    \input{resources/algorithmic/constraints/valid}
    \vspace{-2.2em}
    \input{resources/algorithmic/constraints/bindprem}
  \end{minipage}
  \hfill
  \begin{minipage}[t]{0.48\textwidth}
    \input{resources/algorithmic/constraints/bindexp}
    \vspace{-2.2em}
    \input{resources/algorithmic/constraints/dim}
  \end{minipage}
\vspace*{-1em}
\end{figure}

\begin{figure}[t]
  {\footnotesize
  \input{resources/algorithmic/constraints/il}}
  \vspace{-1em}
  \caption{Abstract syntax of IL}
  \label{fig:il-syntax}
\vspace*{-1em}
\end{figure}

We now discuss the constraints for algorithmic inference rules.
Of the five numbered paragraphs below, paragraphs 1-4 validate the dataflow
property with respect to the function \textsc{Valid} in Alg.~\ref{alg:valid},
and paragraph 5 addresses determinism.
\textsc{Valid} is defined in the IL syntax in
Fig.~\ref{fig:il-syntax}, taking a $\X{rule}$ as input.
It performs dataflow analysis via the \textsc{Bind} functions in
Algs.~\ref{alg:bindprem} and \ref{alg:bindexp}, checking that new $\X{id}$s are
computed only from already known $\X{id}$s (lines 1-7).
Here, $\X{B}$ denotes the set of known $\X{id}$s and $\X{N}$ denotes newly
computed $\X{id}$s.
During the process, side-conditions are inserted to handle partial binding
patterns, as explained in paragraph 1 below.
Then, it annotates the iterators ($\X{iterexp}$ and $\X{iterprem}$) with the
iterated $\X{var}$s, using \textsc{Dim}$_{\textsc{annot}}$ in
Alg.~\ref{alg:dim}, and checks that the iterators are valid with
\textsc{Dim}$_{\textsc{valid}}$ (lines 8-9).
We illustrate the algorithm using the P4 \pfourinl|if| statement typing rule in
Fig.~\ref{fig:p4-spectec}.

\vspace*{-.5em}
\paragraph{1. Relation input annotation}
%
We represent inference rules as a triple of input, premises, and output, making
the dataflow explicit as shown in Fig.~\ref{fig:il-syntax} for the $\X{rule}$
production.
%
%
\vspace*{.5em}
\begin{lstlisting}[
  style=spectecstyle,
  language=spectec,
  basicstyle=\ttfamily\footnotesize,
  backgroundcolor=\color{blue!3},
  aboveskip=0pt,
  belowskip=0pt
]
 @\speckey{rel}@ @\specrel{Stmt\_ok}@: TC |- stmt : TC stmtIR @\speckey{hint}@(input %0 %1)
 @\speckey{rule}@ @\specrel{Stmt\_ok}@/if: TC |- @\specterm{IfS}@ expr stmt_then* stmt_else* : TC stmtIR
\end{lstlisting}
\speckeyinl{rel}s require a \speckeyinl{hint} specifying which
parts of the relation are input.
With this, \specrelinl{Stmt\_ok} is treated as a partial function computing
\specplain{TC} and \specplain{stmtIR} from the input \specplain{TC} and
\specplain{stmt}.
%
%
\textsc{Valid} calls \textsc{Bind}$_{\textsc{Exps}}$ (line~3) to check that the
input \specplain{TC} and \specplain{(}\specterminl{IfS}\specplain{ ...}\specplain{)} of
\specrelinl{Stmt\_ok}\specplain{/if} are valid for binding new
$\X{id}$s.
%
\specplain{TC} trivially binds \specplain{TC}, and
\specplain{(}\specterminl{IfS}\specplain{ ...}\specplain{)} is
a \emph{partial} binding pattern, projecting the input into the
\specterminl{IfS} constructor.
So, it is renamed to a fresh $\X{id}$ (\specplain{stmt\_fresh}),
and a side-condition is inserted to check that $\X{id}$ is a case of
\specterminl{IfS}.
Then, it binds \specplain{expr}, \specplain{stmt\_then*}, and
\specplain{stmt\_else*} from the fresh variable via \K{let}, resulting in:
\vspace*{.5em}
\begin{lstlisting}[
  style=spectecstyle,
  language=spectec,
  basicstyle=\ttfamily\footnotesize,
  backgroundcolor=\color{blue!3},
  aboveskip=0pt,
  belowskip=0pt
]
 @\speckey{rule}@ @\specrel{Stmt\_ok}@/if: TC |- stmt_fresh : TC stmtIR
   -- @\speckey{if}@ stmt_fresh @\speckey{matches}@ @\specterm{IfS}@
   -- @\speckey{let}@ @\specterm{IfS}@ expr stmt_then* stmt_else* = stmt_fresh
\end{lstlisting}
Lines 2-4 handle the case where the expression at a binding site
is already known.
Suppose
\mbox{\specplain{(}\specterminl{IfS}\specplain{ expr stmt\_then*}\specplain{ }\speckeyinl{eps}\specplain{)}}
is an input, where
an empty sequence \speckeyinl{eps} is a bound expression.
\speckeyinl{eps} is replaced with \specplain{stmts\_fresh}, and a
side-condition is inserted to check:
\mbox{\speckeyinl{if}\specplain{ stmts\_fresh = }\speckeyinl{eps}}.

\vspace*{-.5em}
\paragraph{2. Ordered premises}
%
%
%
While premises are unordered in general inference rules, to enforce explicit
dataflow, we require premises to be ordered, from top to bottom.
%
\textsc{Valid} calls \textsc{Bind}$_{\textsc{Prems}}$, with $\X{B}$ derived
from $\X{exp_{in}}^*$ (line~5).
\textsc{Bind}$_{\textsc{Prems}}$ in turn invokes
\textsc{Bind}$_{\textsc{Prem}}$ for each $\X{prem}$ in order, while
incrementally tracking $\X{B}$.
A call to \textsc{Bind}$_{\textsc{Prem}}$ checks that $\X{prem}$ is valid, and
outputs the newly bound $\X{N}$, alongside the inserted side-conditions
$\X{prem_{side}}^*$ and elaborated $\X{prem}$.
The first premise,
\mbox{\specrelinl{Expr\_ok}\specplain{: TC |- expr : exprIR}},
corresponds to line 15 of \textsc{Bind}$_{\textsc{Prem}}$.
It is checked that both \specplain{TC} and \specplain{expr} on the input
of \specrelinl{Expr\_ok} are in $\X{B}$.
The output \specplain{exprIR} is checked with \textsc{Bind}$_{\textsc{Exps}}$,
and added to $\X{N}$.

\vspace*{-.5em}
\paragraph{3. Distinguishing \K{if}-\K{let}}
\K{let} premise is an extension from EL to IL.
\K{let} premises are elaborated from \K{if} premises in EL, to make the
dataflow within \K{if} premises explicit.
\vspace{0.5em}
\noindent
\begin{lstlisting}[
  style=spectecstyle,
  language=spectec,
  basicstyle=\ttfamily\footnotesize,
  backgroundcolor=\color{blue!3},
  aboveskip=0pt,
  belowskip=0pt,
]
  -- @\speckey{if}@ typeIR = @\specfunc{\$typeof}@(exprIR)   ;; becomes @\speckey{let}@ typeIR = @\specfunc{\$typeof}@(exprIR)
  -- @\speckey{if}@ typeIR = @\specterm{BoolT}@             ;; becomes @\speckey{if}@ typeIR = @\specterm{BoolT}@
\end{lstlisting}
\noindent
The above two \K{if} premises serve different purposes.
The first binds \specplain{typeIR} from the known \specplain{exprIR}, while the
second checks whether \specplain{typeIR} equals \specterminl{BoolT}.
The \K{=} is overloaded; it means either a binding or a check.
This ambiguity is resolved by consulting $\X{B}$, as in lines 3-11 of
\textsc{Bind}$_{\textsc{Prem}}$.
If both sides of~\K{=} are known, it is a check and remains an \K{if} premise.
If one side is known and the other is not, it is a binding, and becomes a
\K{let} premise.
Otherwise, it is an error, as the direction of dataflow is ambiguous.
%
%
Using the aforementioned constraints, \textsc{Valid} performs dataflow analysis
over the premises and finally checks that the output $\X{exp_{out}}^*$ contains
only known $\X{id}$s in $\X{B}$ (line 7).

\vspace*{-.5em}
\paragraph{4. Precluding existentials}
Existential meta-variables may appear in three ways in inference rules, and all
of them are disallowed and checked for in the validation algorithm.
Consider the following variants of \mbox{\speckeyinl{let}\specplain{ typeIR = }\specfuncinl{\$typeof}\specplain{(exprIR)}},
binding \specplain{typeIR} from known \specplain{exprIR}:

\vspace{0.5em}
\noindent
\begin{minipage}{0.32\textwidth}
\begin{lstlisting}[
  style=spectecstyle,
  language=spectec,
  basicstyle=\ttfamily\footnotesize,
  backgroundcolor=\color{blue!3},
]
  ;; unknown meta-variable
  -- @\speckey{let}@ typeIR
      = @\specfunc{\$typeof}@(exprIR_unknown)
\end{lstlisting}
\end{minipage}\hfill
\begin{minipage}{0.32\textwidth}
\begin{lstlisting}[
  style=spectecstyle,
  language=spectec,
  basicstyle=\ttfamily\footnotesize,
  backgroundcolor=\color{blue!3},
]
  ;; uncomputable binding
  -- @\speckey{let}@ @\specfunc{\$oftype}@(typeIR)
      = exprIR
\end{lstlisting}
\end{minipage}\hfill
\begin{minipage}{0.32\textwidth}
\begin{lstlisting}[
  style=spectecstyle,
  language=spectec,
  basicstyle=\ttfamily\footnotesize,
  backgroundcolor=\color{blue!3},
]
  ;; unknown dimension
  -- (@\speckey{let}@ typeIR
      = @\specfunc{\$typeof}@(exprIR))*
\end{lstlisting}
\end{minipage}
First, existentials appear as unbound meta-variables, such as
\specplain{exprIR\_unknown} in the first variant.
The analysis checks that all data flowing into newly computed data must
already be known, in \textsc{Bind}$_{\textsc{Prem}}$ for $\X{prem}$ (lines~5,13,16),
and in \textsc{Valid} for $\X{exp_{out}}^*$ (line~7).
Second, existentials may appear within non-invertible binding expressions.
The second variant tries to bind \specplain{typeIR} from the known
\specplain{exprIR}, but there is no algorithm to compute the binding, as it
requires solving the inverse of \specfuncinl{\$oftype}, which is not computable
in general.
Thus, \textsc{Bind}$_{\textsc{Exp}}$ checks that the binding expression is
invertible.
It only allows $\X{id}, \X{term}~\X{exp}^*, \X{exp}^?, \K{[}\X{exp}^*\K{]}$,
and $\X{exp}~\K{::}~\X{exp}$ for binding.
Lastly, existentials may appear as unknown dimensions in iterators.
After binding analysis, \textsc{Valid} calls
\textsc{Dim}$_{\textsc{annot}}$ to annotate the $\X{iterexp}$ and
$\X{iterprem}$ with the iterated $\X{var}$s (line~8).
For instance, a meta-expression \specplain{stmtIR\_then*} is annotated
\specplain{* }\speckeyinl{in}\specplain{ stmtIR\_then*}, meaning that
\specplain{stmtIR\_then*} is iterated into each \specplain{stmtIR\_then}.
$\X{iterprem}$s also have the $\K{out}$ direction, indicating which $\X{var}$s
are yielded by the iteration.
For the third variant, because \specplain{exprIR} is known and is dimension-less,
\textsc{Dim}$_{\textsc{annot}}$ annotates the premise with
\mbox{\specplain{* }\speckeyinl{in}\specplain{ $\epsilon$, }\speckeyinl{out}\specplain{ typeIR*}},
meaning that it yields a list of \specplain{typeIR}s.
However, this premise cannot be computed, as the length of the yielded
\specplain{typeIR*} cannot be determined.
%
%
\textsc{Dim}$_{\textsc{valid}}$, called from line 9 of \textsc{Valid} checks
that the $\X{iterprem}$s contain at least one $\X{var}$ in the \K{in}
direction.


\vspace*{-.5em}
\paragraph{5. Validating determinism}
\speckeyinl{rule}s specify each case of the \speckeyinl{rel}.
If multiple rules can be applied simultaneously,
the semantics may be non-deterministic.
However, this is difficult to check statically, as it requires verifying that
each rule specifies a mutually exclusive case of the input.
This reduces to a non-trivial problem of checking whether two Boolean
meta-expressions are mutually exclusive.
%
Suppose we have two rules with premises
\mbox{\speckeyinl{if}\specplain{ }\specfuncinl{\$odd}\specplain{(n\_a * n\_b)}} and
\mbox{\speckeyinl{if}\specplain{ }\specfuncinl{\$even}\specplain{(n\_a)}\specplain{ \textbackslash}\specplain{/}\specplain{ }\specfuncinl{\$even}\specplain{(n\_b)}},
respectively, where \specplain{n\_a} and \specplain{n\_b} are inputs.
While they are mutually exclusive for all \specplain{n\_a} and
\specplain{n\_b}, statically verifying this is non-trivial.
Hence, it is also challenging to rule out non-determinism that arises from
mistakes, e.g. writing
\mbox{\specfuncinl{\$odd}\specplain{(n\_a + n\_b)}} instead of
\mbox{\specfuncinl{\$odd}\specplain{(n\_a * n\_b)}}.
Instead of statically verifying determinism, we validate it dynamically during
inference rule execution.
For instance, if \specplain{n\_a} is 7 and \specplain{n\_b} is 42, both rules
would be satisfied, reporting a non-determinism error.
%

\subsection{Executing Algorithmic Inference Rules}\label{sec:algorithmic:exec}

Having validated that the inference rules are algorithmic, implementing a
meta-interpreter for them is relatively straightforward.
%
Algorithms for evaluating relations and premises are available in
\cref{app:algorithmic-exec}.
Intuitively, a relation is executed by trying to apply each of its rules in
order.
If a rule application fails, e.g. because its \speckeyinl{if} premise was not
satisfied, we backtrack and try the next rule.
The meta-interpreter offers two modes: a \texttt{-det} mode that validates
determinism by applying all rules and checking that at most one rule is
applicable, and a default mode that applies the first applicable rule.
\texttt{-det} mode is a faithful execution of the algorithmic inference rules,
while the default mode is a lightweight version of it, for performance reasons.

Executing the mechanized P4 static and dynamic semantics yields a P4 type
checker and interpreter backends, respectively.
\specrelinl{Pgm\_ok} is the entry point for the P4 type system specification.
A P4 program is parsed into a meta-value of meta-type \specplain{pgm}.
Within \specplain{pgm}, a P4 \pfourinl|if| statement appears as a meta-value of
meta-type \specplain{stmt}, of case \specterminl{IfS}.
Evaluating \specrelinl{Pgm\_ok} on \specplain{pgm} performs type checking of the
P4 program and produces \specplain{pgmIR} if it is well-typed.
The IL interpreter is written in OCaml and executes the
algorithmic inference rules.
Thus, a P4 program is type checked by running the executable semantics, which
are in turn executed by the IL interpreter in OCaml.
The same applies to the dynamic semantics and the P4 interpreter backend.

\subsection{Extern Interfacing}\label{sec:algorithmic:extern}

\begin{figure}[t]
  \begin{minipage}[t]{\textwidth}
\begin{lstlisting}[
      style=spectecstyle,
      language=spectec,
      basicstyle=\ttfamily\footnotesize,
      backgroundcolor=\color{blue!3},
    ]
@\emptyline@
 @\speckey{rule}@ @\specrel{Stmt\_eval}@/ret: STO EC |- @\specterm{RetS}@ @\speckey{eps}@ : STO EC @\specterm{RET}@
@\emptyline@
 @\speckey{rule}@ @\specrel{Stmt\_eval}@/call: STO_0 EC |- @\specterm{CallS}@ id id_method typeIR* exprIR* : STO_1 EC @\specterm{CONT}@
   -- @\speckey{if}@ @\specterm{ExternO}@ state = @\specfunc{\$find\_sto}@(STO_0, id)
   -- (@\specrel{Expr\_eval}@: EC |- exprIR : val)*
   -- @\specrel{Call\_ext}@: state |- id id_method typeIR* val* : state_new
   -- @\speckey{if}@ obj_new = @\specterm{ExternO}@ state_new
   -- @\speckey{if}@ STO_1 = @\specfunc{\$update\_sto}@(STO_0, id, obj_new)
@\emptyline@
 @\speckey{rule}@ @\specrel{Stmts\_eval}@/nil: STO EC |- @\speckey{eps}@ : STO EC @\specterm{CONT}@
@\emptyline@
 @\speckey{rule}@ @\specrel{Stmts\_eval}@/ret: STO_0 EC_0 |- stmtIR_h :: stmtIR_t* : STO_1 EC_1 @\specterm{RET}@
   -- @\specrel{Stmt\_eval}@: STO_0 EC_0 |- stmtIR_h : STO_1 EC_1 @\specterm{RET}@
@\emptyline@
 @\speckey{rule}@ @\specrel{Stmts\_eval}@/cont: STO_0 EC_0 |- stmtIR_h :: stmtIR_t* : STO_2 EC_2 sign
   -- @\specrel{Stmt\_eval}@: STO_0 EC_0 |- stmtIR_h : STO_1 EC_1 @\specterm{CONT}@
   -- @\specrel{Stmts\_eval}@: STO_1 EC_1 |- stmtIR_t* : STO_2 EC_2 sign
\end{lstlisting}
    \vspace{-1em}
    \caption{Evaluation rules for P4 statements.}
    \label{fig:stmts-rules}
  \end{minipage}
    \vspace{-1em}
\end{figure}

The P4 dynamic semantics is \emph{open-world}.
A P4 program only specifies the behavior of programmable blocks within the
architecture that the program is written for.
The architecture is responsible for invoking the programmable blocks in the
intended order.
Furthermore, the behavior of \pfourinl|extern|s is not specified by the P4
program, and must be provided by the architecture as well.
Hence, to execute the end-to-end packet processing pipeline, the IL interpreter
running the P4 semantics and the architecture must cooperate.
Referring back to the VSS architecture in Fig.~\ref{fig:background:p4:arch}, the IL
interpreter runs the \textbf{\texttt{\small Parser}},
\textbf{\texttt{\small Match-action(s)}}, and \textbf{\texttt{\small Deparser}} blocks, while
VSS calls these blocks in order (denoted by the arrows) and runs the
\textbf{\texttt{\small Input/Output Engine}}.
On each packet entry, the architecture requests the IL interpreter to run
\mbox{\speckeyinl{rel}\specplain{ }\specrelinl{Call\_p4}} in
Fig.~\ref{fig:p4-eval-rel} to execute the parser block.
%
%
Once block execution finishes, the architecture calls the remaining blocks to
complete the pipeline.
During evaluation, \pfourinl|extern| calls may be encountered,
such as \pfourinl|p.emit(hdr)| emitting \specplain{hdr} to the network.
%
%
%
%
In Fig.~\ref{fig:stmts-rules}, the \specplain{call} rule invokes
\mbox{\speckeyinl{ext rel}\specplain{ }\specrelinl{Call\_ext}} at the third
premise,
which invokes the corresponding extern implementation provided by the
architecture and returns the result to the IL interpreter.

Architectures differ in their pipeline stages and the set of supported externs.
Therefore, architecture implementations are \emph{plugged in} to P4-SpecTec,
allowing to execute packet processing for different architectures while
evaluating programmable blocks using the mechanized semantics.

A problem arises when backtracking of inference rules interacts
with the side-effects of externs.
%
%
Suppose we backtrack the rules in Fig.~\ref{fig:stmts-rules} on the
following code with side-effecting \pfourinl|p.emit(hdr)|:
\begin{lstlisting}[
       style=p4style,
       language=p4,
       basicstyle=\ttfamily\footnotesize,
     ]
if (hdr.src != hdr.dst) p.emit(hdr); return;
\end{lstlisting}
%
Suppose \pfourinl|hdr.src| and \pfourinl|hdr.dst| are distinct.
\speckeyinl{Stmts\_eval}\specplain{/nil} fails because the statement sequence
is not empty.
Then we try the \specplain{ret} rule.
Its first premise evaluates the \pfourinl|if| statement, performing the
\pfourinl|p.emit(hdr)| call and resulting in \specplain{sign\_h} being
\specterminl{CONT}.
Because it does not equal \specterminl{RET}, the rule fails and we backtrack to
try the \specplain{cont} rule.
We again evaluate the first \pfourinl|if| statement, triggering
\pfourinl|p.emit(hdr)| again.
Thus, the same extern call is performed twice, emitting the same header twice.

To solve this problem, P4-SpecTec encodes the external state as
\speckeyinl{ext syn}~\specplain{state},
%
%
where \specplain{state} captures the internal state of the externs, such as the state
of the header being assembled in the deparser via \pfourinl|emit| calls.
Within the inference rules, \specplain{state} is an opaque value.
%
On the extern implementation side, \specplain{state} is implemented as a JSON
object.
On \speckeyinl{ext rel} calls, the current \specplain{state} is passed as input
to the architecture, deserialized into the internal state, updated by the
extern implementation, and serialized back to JSON before returning to the IL
interpreter.
For \pfourinl|emit|s, its internal state is the header content being assembled.
With this design, the side-effects are controlled, yet the internal details of
the externs are not exposed to the mechanized specification.


\section{Structured Inference Rules}\label{sec:structured}

\begin{figure}[t]
  {\footnotesize
  \input{resources/structured/sl}}
  \vspace{-1.3em}
  \caption{Abstract syntax of SL}
  \label{fig:sl-syntax}
  \vspace{-1.4em}
\end{figure}


While algorithmic inference rules make language semantics executable, they
carry redundant information.
%
In Fig.~\ref{fig:stmts-rules}, \mbox{\specrelinl{Stmts\_eval}\specplain{/ret}}
and \specplain{cont} overlap in their first premises, both evaluating
\specplain{stmtIR\_h}.
%
Hence, if \specplain{stmtIR\_h} does not \specterminl{RET}, the same statement
is evaluated twice.
Similarly for the \mbox{\specrelinl{Stmt\_eval}\specplain{/if-then}} and
\specplain{if-else} rules in Fig.~\ref{fig:p4-eval-rule}, the condition
expression is evaluated twice if it evaluates to
\mbox{\specterminl{BoolV}\specplain{ }\speckeyinl{false}}.
Redundancy propagates across relation invocations.
If \specplain{stmtIR\_h} is an \pfourinl|if| statement with a false condition that does
not return, the condition gets evaluated four times.
%

Additionally, inference rules lack structure.
%
%
%
For specification readers, the semantics of a language construct is spread
across multiple inference rules, making it difficult to follow the control flow
of the specification.
Rather than presenting two separate rules for \specplain{if-then} and
\specplain{if-else}, it is more intuitive to present them as a single algorithm
with an explicit branching on the condition expression.

To address these, the mechanized specification in IL is processed to \emph{SL
(Structured Language)} via a process we call \emph{structuring}, where
redundant information is removed and structured control flow is introduced.
See Fig.~\ref{fig:sl-syntax} for the abstract syntax of SL.
From the definition of $\X{rel}$, notice that $\X{rule}$s in IL are merged into
a sequence of $\X{instr}$s operating on the same input $\X{exp}^*$.
An $\X{instr}$ is similar to a $\X{prem}$ in IL.
Yet, the iterator of an iterated premise is inlined into each $\X{instr}$ as
$\X{iterexp}$ in $\K{if}$ and $\X{iterinstr}$ in $\K{let}$ and relation
invocation.
An $\K{if}$ instruction has an optional $\K{else}$ branch, representing a
structured control flow construct.
The output part of a relation is represented as a $\K{result}$ instruction.

The transformation from IL to SL consists of the following steps: (i) merging,
(ii) conversion into instructions (iii) removing redundancy and (iv) inserting
$\K{else}$ branches.
We illustrate these steps with the
\mbox{\specrelinl{Stmt\_eval}\specplain{/if-then}} and \specplain{if-else}
rules.
The input part of the two rules are elaborated as below, with side conditions
inserted for partial bindings.

\vspace{0.5em}
\noindent
\begin{minipage}{0.48\textwidth}
\centering
\begin{lstlisting}[
  style=spectecstyle,
  language=spectec,
  basicstyle=\ttfamily\footnotesize,
  backgroundcolor=\color{blue!3},
]
 @\speckey{rule}@ @\specrel{Stmt\_eval}@/if-then:
  STO_0 EC_0 |- stmtIR_fresh_0 : ...
  -- @\speckey{if}@ stmtIR_fresh_0 @\speckey{matches}@ @\specterm{IfS}@
  -- @\speckey{let}@ @\specterm{IfS}@ exprIR stmtIR_then* stmtIR_else*
      = stmtIR_fresh_0
\end{lstlisting}
\end{minipage}
\hfill
\begin{minipage}{0.48\textwidth}
\centering
\begin{lstlisting}[
  style=spectecstyle,
  language=spectec,
  basicstyle=\ttfamily\footnotesize,
  backgroundcolor=\color{blue!3},
]
 @\speckey{rule}@ @\specrel{Stmt\_eval}@/if-else:
  STO_0 EC_0 |- stmtIR_fresh_1 : ...
  -- @\speckey{if}@ stmtIR_fresh_1 @\speckey{matches}@ @\specterm{IfS}@
  -- @\speckey{let}@ @\specterm{IfS}@ exprIR stmtIR_then* stmtIR_else*
      = stmtIR_fresh_1
\end{lstlisting}
\end{minipage}
%
%
The two rules have the same \specplain{STO\_0} and \specplain{EC\_0}
inputs, but differ in \specplain{stmtIR\_fresh\_0} and
\specplain{stmtIR\_fresh\_1}.
To safely merge them, we apply \emph{anti-unification}~\cite{antiunify} to the
inputs of the rules, which computes the most specific generalization of the
inputs.
Anti-unification produces a common input, \specplain{stmtIR\_au}.
%

\vspace{0.5em}
\noindent
\begin{minipage}{0.48\textwidth}
\centering
\begin{lstlisting}[
  style=spectecstyle,
  language=spectec,
  basicstyle=\ttfamily\footnotesize,
  backgroundcolor=\color{blue!3},
]
 @\speckey{rule}@ @\specrel{Stmt\_eval}@/if-then:
  STO_0 EC_0 |- stmtIR_au : ...
  -- @\speckey{let}@ stmtIR_fresh_0 = stmtIR_au
\end{lstlisting}
\end{minipage}
\hfill
\begin{minipage}{0.48\textwidth}
\centering
\begin{lstlisting}[
  style=spectecstyle,
  language=spectec,
  basicstyle=\ttfamily\footnotesize,
  backgroundcolor=\color{blue!3},
]
 @\speckey{rule}@ @\specrel{Stmt\_eval}@/if-else:
  STO_0 EC_0 |- stmtIR_au : ...
  -- @\speckey{let}@ stmtIR_fresh_1 = stmtIR_au
\end{lstlisting}
\end{minipage}
\noindent Additional $\K{let}$ premises are added to bind
\specplain{stmtIR\_fresh\_0} and \specplain{stmtIR\_fresh\_1} from
\specplain{stmtIR\_au}.
Once the rule inputs are merged, premises and the output of each rule
are converted into instructions as:
%
%

\vspace{0.5em}
\noindent
\begin{lstlisting}[
  style=spectecstyle,
  language=spectec,
  basicstyle=\ttfamily\footnotesize,
  backgroundcolor=\color{blue!3},
  aboveskip=0pt,
  belowskip=0pt
]
 @\speckey{rel}@ @\specrel{Stmt\_eval}@(STO_0, EC_0, stmtIR_au):
  1. @\speckey{let}@ stmtIR_fresh_0 = stmtIR_au @\speckey{in}@
    a. @\speckey{if}@ stmtIR_fresh_0 @\speckey{matches}@ @\specterm{IfS}@ @\speckey{then}@
      i. @\speckey{let}@ @\specterm{IfS}@ exprIR stmtIR_then* stmtIR_else* = stmtIR_fresh_0 @\speckey{in}@ ...
        x. @\speckey{result in}@ STO_1, EC_1, sign
  2. @\speckey{let}@ stmtIR_fresh_1 = stmtIR_au @\speckey{in}@
    a. @\speckey{if}@ stmtIR_fresh_1 @\speckey{matches}@ @\specterm{IfS}@ @\speckey{then}@
      i. @\speckey{let}@ @\specterm{IfS}@ exprIR stmtIR_then* stmtIR_else* = stmtIR_fresh_1 @\speckey{in}@ ...
        y. @\speckey{result in}@ STO_1, EC_1, sign
\end{lstlisting}
An $\K{if}$ premise becomes an $\K{if}$ instruction without an $\K{else}$
branch.
A $\K{let}$ premise becomes a $\K{let}$ instruction with subsequent premises
converted into instructions and nested into the $\K{let}$ body.
Similar conversion applies to the relation invocation premises.
Finally, the output expressions are converted into a $\K{result}$ instruction.
Then, instructions from each rule are concatenated, where the backtracking
evaluation of the IL interpreter is made explicit.

Redundancy is present in the above instruction sequence.
\specplain{stmtIR\_fresh\_0} and \specplain{stmtIR\_fresh\_1} are both bound to
\specplain{stmtIR\_au}.
Also, instructions 1.a and 2.a are essentially the same, as they both check
if \specplain{stmtIR\_au} matches \specterminl{IfS}.
Redundant bindings and instructions are removed, thus yielding:
\vspace*{0.5em}
\begin{lstlisting}[
  style=spectecstyle,
  language=spectec,
  basicstyle=\ttfamily\footnotesize,
  backgroundcolor=\color{blue!3},
  aboveskip=0pt,
  belowskip=0pt
]
  1. @\speckey{if}@ stmtIR_au @\speckey{matches}@ @\specterm{IfS}@ @\speckey{then}@
    a. @\speckey{let}@ @\specterm{IfS}@ exprIR stmtIR_then* stmtIR_else* = stmtIR_au @\speckey{in}@ ...
      x.  @\speckey{if}@ val @\speckey{=}@ @\specterm{BoolV}@ @\speckey{true}@ @\speckey{then}@ ...
      xi. @\speckey{if}@ val @\speckey{=}@ @\specterm{BoolV}@ @\speckey{false}@ @\speckey{then}@ ...
\end{lstlisting}
\noindent Removal is safe as the specification is side-effect-free.
No global meta-variable exists in the meta-language.
Also, external side-effects are managed by \speckeyinl{ext syn}, as explained
in Section~\ref{sec:algorithmic:extern}.
%

Finally, structured control flow is injected as $\K{else}$ branches of $\K{if}$
instructions.
Instructions 1.a.x and 1.a.xi representing mutually exclusive branches are
structured into an $\K{if-then-else}$ construct:
\vspace*{.5em}
\begin{lstlisting}[
  style=spectecstyle,
  language=spectec,
  basicstyle=\ttfamily\footnotesize,
  backgroundcolor=\color{blue!3},
  aboveskip=0pt,
  belowskip=0pt
]
      x.  @\speckey{if}@ val @\speckey{=}@ @\specterm{BoolV}@ @\speckey{true}@ @\speckey{then}@ ...
      xi. @\speckey{else if}@ val @\speckey{=}@ @\specterm{BoolV}@ @\speckey{false}@ @\speckey{then}@ ...
\end{lstlisting}
Consecutive \K{if} instructions are examined for structuring opportunities.
Using heuristics, we identify \K{if} conditions with a common structure,
\mbox{\specplain{(}\speckeyinl{if}\specplain{ val = ...)}} in the above
example.
Then, we apply some pattern-specific rules checking whether the candidates are
mutually exclusive.
Because
\mbox{\specterminl{BoolV}\specplain{ }\speckeyinl{true}} and
\mbox{\specterminl{BoolV}\specplain{ }\speckeyinl{false}} are distinct,
the rules confirm that the candidates are mutually exclusive
and structure them as an \K{if-then-else}.
This may miss some structuring opportunities, but as premises in the mechanized
specification follow structured patterns, it is sufficient to capture most
opportunities.

After structuring, executing the SL specification is straightforward.
%
%
The SL interpreter also supports the \texttt{-det} flag to dynamically check
determinism.
The semantics preservation of structuring is validated by tests, where IL and
SL executions show the same results (\cref{sec:eval:structuring}).

We summarize the EL, IL, and SL stages in the P4-SpecTec architecture.
Parsing the mechanized specification yields EL.
We then elaborate EL to IL, checking that the mechanized inference rules
are algorithmic.
General inference rules are declarative, where their semantics is essentially a
proof search, which is non-algorithmic, as it may incorporate existential
variables and non-determinism.
Elaboration restricts them to an algorithmic subset, where dataflow is explicit
and determinism is dynamically validated.
The IL interpreter performs backtracking on inference rules.
Then via structuring, IL is converted to SL, where redundant backtracking is
identified and removed.

\section{Generating the Prose Specification}\label{sec:prose}

SL is finally compiled to \emph{PL (Prose Language)} for the specification
document backend.
PL is
the same in structure as SL, but with prose explanations inserted.
PL is printed in AsciiDoc format, which the P4 official specification uses.
AsciiDoc can then be rendered as HTML or PDF as shown in
Fig.~\ref{fig:p4-spectec-arch} (b).
%
%
While the compilation to PL is mostly straightforward, two extensions allow the
generation of more readable prose specifications: prose \speckeyinl{hint}s and
\speckeyinl{rulegroup}s.

\begin{lstlisting}[
  style=spectecstyle,
  language=spectec,
  basicstyle=\ttfamily\footnotesize,
  backgroundcolor=\color{blue!3},
]
 @\speckey{rel}@ @\specrel{Stmts\_eval}@: STO EC |- stmtIR* : STO EC sign
    @\speckey{hint}@(input %0 %1 %2)
    @\speckey{hint}@(prose_in "evaluating" %2 "under context" %1 "and store" %0)
    @\speckey{hint}@(prose_out "store" %3 "," context" %4 "and" %5)
\end{lstlisting}

Prose \speckeyinl{hint}s can be annotated alongside \speckeyinl{rel}
declarations.
Above, the \specplain{prose\_in} hint specifies the prose explanation for the
input of \specplain{Stmts\_eval} and
%
%
\specplain{prose\_out} specifies its output.
During compilation to PL, the hints are inlined into the SL instructions.
Thus, for the following premise:
\begin{lstlisting}[
  style=spectecstyle,
  language=spectec,
  basicstyle=\ttfamily\footnotesize,
  backgroundcolor=\color{blue!3},
]
 -- @\specrel{Stmts\_eval}@: STO_0 EC_0 |- stmtIR_then* : STO_1 EC_1 sign
\end{lstlisting}
\noindent hints are inlined and printed in AsciiDoc as:
\begin{lstlisting}[
  basicstyle=\ttfamily\scriptsize,
]
Let store `STO~1~`, context `EC~1~`, and `sign` be the result of
xref:Stmts_eval[evaluating `stmtIR~then~^{asterisk}^` under context `EC~0~` and store `STO~0~`].
\end{lstlisting}
\noindent When rendered, it becomes step 2.a shown in Fig.~\ref{fig:p4-spectec-arch} (b).
Notice that the cross-reference link {\small\texttt{xref:Stmts\_eval}} is also
inserted, which allows readers to easily jump to the definition of
\specrelinl{Stmts\_eval}.

Another extension is the \speckeyinl{rulegroup} declaration.
After the IL-to-SL transformation, all rules in a relation are merged into a
\emph{single} instruction sequence,
which may reduce readability.
%
%
%
For instance, in the \pfourinl|if| section of the P4 specification document, we
only want to show the prose algorithm for
\mbox{\specrelinl{Stmt\_eval}\specplain{/if-then}} and \specplain{if-else}
rules, but not the \specplain{call} and \specplain{ret} rules in
Fig.~\ref{fig:stmts-rules}.
%
%
\begin{lstlisting}[
  style=spectecstyle,
  language=spectec,
  basicstyle=\ttfamily\footnotesize,
  backgroundcolor=\color{blue!3},
]
 @\speckey{rulegroup}@ @\specrel{Stmt\_eval}@/if
 { @\speckey{rule}@ @\specrel{Stmt\_eval}@/if-then: STO_0 EC_0 |- @\specterm{IfS}@ exprIR stmtIR_then* stmtIR_else* : ...
   @\speckey{rule}@ @\specrel{Stmt\_eval}@/if-else: STO_0 EC_0 |- @\specterm{IfS}@ exprIR stmtIR_then* stmtIR_else* : ... }
\end{lstlisting}
\noindent The above \speckeyinl{rulegroup} specifies that \specplain{if-then}
and \specplain{if-else} rules belong to the \specplain{if} group.
%
It is rendered as Fig.~\ref{fig:p4-spectec-arch} (b).
Notice that its title shows \mbox{\specplain{(}\specterminl{IfS}\specplain{ ...)}}
instead of \specplain{stmtIR\_au}, which appeared after inserting
side-conditions for partial binds (\cref{sec:algorithmic-constraints})
and anti-unification (\cref{sec:structured}).
To preserve the syntax-directed presentation, the inputs of the rules in a
\speckeyinl{rulegroup} are first anti-unified prior to side-condition
insertion.
This yields \specplain{STO\_0}, \specplain{EC\_0}, and
\mbox{\specterminl{IfS}\specplain{ exprIR stmtIR\_then* stmtIR\_else*}}
for the rules in the \specplain{if} group.
Side-conditions for partial binds are then inserted for the anti-unified term,
but hidden in the prose specification.
The hidden side-conditions are still used for the IL and SL interpreters.
%

Finally, the PL specification is rendered as a specification document.
Although the current official P4 specification is informal, we do \emph{not}
intend to replace everything.
The current specification is well-written, conveying high-level intuition,
which is not captured in the \speckeyinl{rule}s themselves.
To this end, we \emph{splice}, or insert, the PL specification into the current
document, as done in Wasm-SpecTec.

\begin{lstlisting}[
  basicstyle=\ttfamily\footnotesize,
  escapeinside={(*}{*)}
]
  [...] the condition expression in P4 is required to be a Boolean [...].
  (*\color{slateblue}\textbf{\$\{rulegroup-prose: Stmt\_eval/if\}}*)
\end{lstlisting}
\noindent The source AsciiDoc document is written as a mixture of natural
language and the \emph{splice anchors} used for inserting the PL specification.
P4-SpecTec inserts the generated AsciiDoc snippet for the anchor
\speckeyinl{\$\{rulegroup-prose: Stmt\_eval/if\}},
which is then rendered as the specification document.

\section{Integration into the P4 Ecosystem}\label{sec:integration}

As of March 2026, P4-SpecTec is \emph{conditionally
adopted}~\cite{p4spectecacc} by the P4 LDWG as the official P4 specification
authoring toolchain.
%
%
Under this model, discussions within LDWG are made at the meta-language level.
Specification edits are cross-validated with the p4c test suite, giving
confidence in updates.
Once the updates are accepted, the specification document is automatically
generated.
Achieving this milestone extends beyond technical challenges.
It also requires socio-technical practices for presenting, persuading, and
introducing changes to an industry-scale language specification.
Below, we share our experience and lessons learned in this process.

\vspace*{-.5em}
\paragraph{Understand the Governance}
The first step
is to understand the governance of the language.
%
P4 is an open project hosted by the Linux Foundation.
The P4 Technical Steering Team (TST)~\cite{p4tst} oversees the project and its technical
direction.
Under the TST, several working groups exist, including the LDWG.
%
The group is open to public participation
%
and the LDWG chair facilitates discussions and decisions.
%
We contacted the LDWG chair in February 2024 and
presented our vision for P4-SpecTec to the broader P4 community in October
2024~\cite{p4spectec2024}.

\vspace*{-.5em}
\paragraph{Engage Continuously}
%
%
We actively engaged with the LDWG in understanding the P4 language.
P4-SpecTec was discussed in 11 monthly LDWG meetings, where we communicated our
vision, shared our progress, collected feedback, and clarified the P4
semantics.
Based on the understanding we gained from the LDWG, we co-designed the
meta-language semantics and the mechanized P4 specification.
%
%
Engagement not only clarified the P4 semantics, but also improved our
understanding of the community's needs, which informed our design decisions.
%
%
Because the P4 community has a network background rather than PL,
we decided to use algorithmic inference rules that are directly executable,
rather than declarative specifications that abstract away execution details.
%

\vspace*{-.5em}
\paragraph{Demonstrate Practical Value}
A practical value of P4-SpecTec that the LDWG appreciated was cross-validation
of the mechanized specification against the p4c test suite.
This
revealed inconsistencies between the specification and p4c.
%
%
Some cases even revealed \emph{reasonable} divergence for better usability in
practice, such as relaxing constraints on implicit casts that the specification
originally disallowed.
In such cases, we amended the official specification to better reflect
real-world practice through LDWG discussions.
%
%
In one discussion, an LDWG member commented: \emph{``Adding the new entries
[...] looks reasonable to me. Thanks for the links to relevant test cases in
the existing p4c test suite.''}
\footnote{\url{https://github.com/p4lang/p4-spec/issues/1324}}
Over two years, we found \nBug{} inconsistencies in the specification and p4c
(\cref{sec:eval:bugs}).

\vspace*{-.5em}
\paragraph{Transition Incrementally}
Integrating a new toolchain into an established ecosystem should not be
intrusive.
The primary tangible change introduced by P4-SpecTec is the auto-generated
specification document.
We therefore adopted an incremental transition approach, gradually inserting
splice anchors to the current specification while keeping the rest of the
document unchanged.

\vspace*{-.5em}
\paragraph{Clarify the Scope}
The mechanization is not free from errors and it does not cover all aspects of
the P4 language.
However, this does not diminish the value of mechanization.
For example, for adopting ESMeta into the JavaScript specification workflow,
Michael Ficarra, one of the specification editors, commented: \emph{``ESMeta
doesn't cover modules and atomics, but it's okay, it doesn't have to cover
everything to be useful.''}
\footnote{\url{https://pldi25.sigplan.org/details/rpls-2025-papers/7/The-Software-Supporting-the-JavaScript-Language-Specification}}
We clarified and presented the scope of P4-SpecTec to the LDWG.

\vspace*{-.5em}
\paragraph{Propose a Clear Roadmap}
Integration is a long-term process requiring continuous effort and commitment
from both the mechanization team and the community.
%
%
%
%
%
After we have gained confidence in P4-SpecTec by validation with the p4c test
suite, we initiated a LDWG discussion to decide whether to adopt P4-SpecTec in
March 2026.
We presented our vision, the status of P4-SpecTec in terms of validation
results, the scope of P4-SpecTec, and a year-long roadmap for integration.
The roadmap included: (i) \emph{documentation} of the meta-language and
toolchain, (ii) a \emph{tutorial} illustrating how to add new features to P4,
and (iii) \emph{integration} into the P4 specification repository with a CI
workflow.
LDWG members raised concerns about the usability of P4-SpecTec, which we
promised in (i) and (ii).
In addition, LDWG members added, (iv) having a transitional period where the
specification is both written in the current manual version and in
P4-SpecTec, to address long-term maintenance concerns.
Conditioned on addressing (i)-(iv), P4-SpecTec was conditionally adopted.

\smallskip
We believe these lessons provide insights for integrating mechanization into
other real-world languages.
Aside from the socio-technical aspects, we believe large parts of P4-SpecTec
can be reused for mechanizing other languages.
The P4-SpecTec architecture, from EL to PL, is not specific to P4, aside from
the \speckeyinl{ext} construct for P4's open-world semantics.
If a language semantics can be specified with algorithmic inference rules,
meta-interpretation and prose generation can be applied.
Applying the P4-SpecTec framework and the socio-technical lessons to
integrating mechanization into other languages is left as future work.


\section{Evaluation}\label{sec:eval}

We evaluate P4-SpecTec with the following research questions:

\begin{itemize}
  \item [\textbf{RQ1}.]
    \textbf{Adequacy of Algorithmic Inference Rules}\vspace{0.1em}\\
    Can algorithmic inference rules specify, execute, and document the P4
    semantics? (\cref{sec:eval:algorithmic})
  \item [\textbf{RQ2}.]
    \textbf{Effectiveness of Structuring}\vspace{0.1em}\\
    Does structuring effectively remove redundancy and insert control flow?
    (\cref{sec:eval:structuring})
  \item [\textbf{RQ3}.]
    \textbf{Comparison with Existing Formalizations}\vspace{0.1em}\\
    Does P4-SpecTec cover more P4 features than existing P4 formalization
    efforts? (\cref{sec:eval:compare})
  \item [\textbf{RQ4}.]
    \textbf{Bugs Found}\vspace{0.1em}\\
    Can specification mechanization reveal bugs in the P4 language ecosystem?
    (\cref{sec:eval:bugs})
\end{itemize}

\subsection{Implementation and Experimental Setup}\label{sec:eval:setup}

P4-SpecTec is implemented in OCaml, reusing the Wasm-SpecTec
frontend for parsing the meta-language.
We removed Wasm-specific features and backends and implemented validation and
execution of algorithmic inference rules for P4.
The mechanization targets the latest version of P4, v1.2.5, covering the P4
concrete syntax, type system, instantiation rules, and dynamic semantics.
Overall, it comprises \locspec{} lines, including \nSyn{}
\speckeyinl{syn}s, \nRel{} \speckeyinl{rel}s, \nRule{}
\speckeyinl{rule}s, and \nDec{} \speckeyinl{dec}s.
%

To validate the executable backends, we use the test suite from the p4c
compiler repository~\cite{p4crepo}.
We use its latest version, v1.2.5.10, released in December 2025.
For the type checker backend,
we use \nPOS{} positive (\POS{}) and \nNEG{} negative (\NEG{}) type
checker tests, located in the {\small\texttt{p4\_16\_samples}} and
{\small\texttt{p4\_16\_errors}} subdirectories of {\small\texttt{testdata}},
respectively.
%
%
We adapt the Petr4 parser to parse P4 programs into meta-values,
which we then pass as input to the typing and instantiation relations.
%

To evaluate the P4 interpreter backend, inputs to P4 programs are required.
These are provided as STF test files, also from p4c.
For example, the STF file in Fig.~\ref{fig:stf} tests \progname{switch.p4} in
Fig.~\ref{fig:background:p4:switch}.
\textbf{\small\texttt{add}} command adds a table entry to
\pfourinl|table access_control|, \textbf{\small\texttt{packet}} inputs a packet at port
{\small\texttt{42}} with header {\small\texttt{0xAABB}}, and
\textbf{\small\texttt{expect}} expects the packet to be emitted at port
{\small\texttt{1}} with {\small\texttt{0xAABB}}.
These STF tests are paired with P4 programs.
%
Recall that the P4 semantics is \emph{open-world}, depending on the external
semantics of the architecture.
To model this, we implement simulators for three widely used P4 architectures:
V1Model, eBPF, and PSA.
Simulators are implemented in OCaml and plugged into the P4 interpreter
backend, as explained in Section~\ref{sec:algorithmic:extern}.
In {\small\texttt{p4\_16\_samples}}, there are \nVMOD{}, \nEBPF{}, and \nPSA{}
P4-STF pairs for the three architectures, which we label \VMOD{}, \EBPF{}, and
\PSA{}, respectively.

We augment the P4-STF pairs using P4Testgen~\cite{p4testgen}, an STF test
generator for P4 programs residing in the p4c repository.
%
%
Using this tool, we generate up to 10 STF tests per P4 program.
This yields \nVMODPlus{} and \nEBPFPlus{} additional test pairs for V1Model
and eBPF, which we denote as \VMODPlus{} and \EBPFPlus{}, respectively.
PSA is unsupported by P4Testgen, so no additional tests are generated for it.

\begin{table}[t]
\centering
\caption{Categorization of excluded tests in \textsc{Future}, \textsc{P4Spec},
\textsc{P4C}, and \textsc{Specific}. \texttt{St} and \texttt{Dy} indicate
the number of excluded static and dynamic tests, respectively. \textbf{CB}
references the confirmed compiler bugs in Table~\ref{tab:eval:bugs}.}
\label{tab:eval:excluded}
\vspace{-0.6em}
\renewcommand{\arraystretch}{1.0}
\setlength{\tabcolsep}{2pt}
\newcommand{\catcell}[2]{\multirow{#1}{*}{\rotatebox[origin=c]{90}{#2}}}
\newcolumntype{P}{>{\raggedright\arraybackslash}p{0.80\linewidth}} 
\newcolumntype{Q}{>{\raggedright\arraybackslash}p{0.84\linewidth}} 
%
\begin{minipage}[t]{0.43\linewidth}
\vspace{0pt}
\centering
{\scriptsize
\begin{tabular}{cPc@{\hspace{4pt}}c@{}}
\toprule
& \textbf{Sub-category} & \texttt{St} & \texttt{Dy} \\ \midrule
\catcell{4}{\textsc{Future}}
 & \texttt{+:} slice operator              & 3 & -- \\
 & Arrays of non-header type                      & 7 & -- \\
 & Compound assignment (e.g., \texttt{+=}) & 6 & -- \\
 & String concatenation                           & 2 & -- \\ \midrule
 & \textbf{Total}                                 & \textbf{18} & \textbf{0} \\ \bottomrule
\end{tabular}}

\vspace{0.3em}
{\scriptsize
\begin{tabular}{cPc@{\hspace{4pt}}c@{}}
\toprule
& \textbf{Sub-category} & \texttt{St} & \texttt{Dy} \\ \midrule
\catcell{21}{\textsc{P4C}}
 & Refer to \textbf{CB2}                                                  & 1 & -- \\
 & Refer to \textbf{CB3}                                                  & 6 & -- \\
 & Refer to \textbf{CB4}                                                  & 3 & -- \\
 & Refer to \textbf{CB5}                                                  & 1 & -- \\
 & Refer to \textbf{CB6}                                                  & 3 & -- \\
 & Refer to \textbf{CB7}                                                  & 1 & -- \\
 & Refer to \textbf{CB8}                                                  & 1 & -- \\
 & Refer to \textbf{CB9}                                                  & 5 & -- \\
 & Refer to \textbf{CB10}                                                 & 1 & -- \\
 & Refer to \textbf{CB11}                                                 & 1 & -- \\
 & Refer to \textbf{CB12}                                                 & -- & 3 \\
 & Refer to \textbf{CB13}                                                 & -- & 3 \\
 & Refer to \textbf{CB14}                                                 & -- & 2 \\
 & Refer to \textbf{CB15}                                                 & -- & 1 \\
 & Tuple fields erroneously treated as l-values                           & 1 & -- \\
 & Wrong scope on abstract method initialization                          & 2 & -- \\
 & Missing support for nested type specializations                       & 1 & -- \\
 & Missing support for a self-cast to a \pfourinlsmall|struct|            & 1 & -- \\
 & Fail to reject cast from \pfourinlsmall|int| to serializable enum      & 1 & -- \\
 & Fail to reject \pfourinlsmall|ternary| match on \pfourinlsmall|error|  & 1 & -- \\
 & Incorrect expected output in a STF test                                & -- & 1 \\ \midrule
 & \textbf{Total}                                                         & \textbf{30} & \textbf{10} \\ \bottomrule
\end{tabular}}
\end{minipage}
\hfill
\begin{minipage}[t]{0.55\linewidth}
\vspace{0pt}
\centering
{\scriptsize
\begin{tabular}{cQc@{\hspace{4pt}}c@{}}
\toprule
& \textbf{Sub-category} & \texttt{St} & \texttt{Dy} \\ \midrule
\catcell{9}{\textsc{P4Spec}}
 & Division/modulo on fixed-width integers                                        & 9 & -- \\
 & Matching control types with default parameters                                 & 1 & -- \\
 & Implicit type constraints imposed by default parameters                        & 1 & -- \\
 & Scope of control apply parameters                                              & 2 & -- \\
 & Meaning of assignments in the form of \pfourinlsmall|X a = a|                  & 1 & -- \\
 & \pfourinlsmall|@optional| annotation on abstract method parameters             & 2 & -- \\
 & Indeterminate type inference result                                            & 5 & -- \\
 & Checking equivalence of table default action parameters                        & 2 & -- \\
 & \pfourinlsmall|ternary| match on \pfourinlsmall|bool|                          & 1 & -- \\ \midrule
 & \textbf{Total}                                                                 & \textbf{24} & \textbf{0} \\ \bottomrule
\end{tabular}}

\vspace{0.3em}
{\scriptsize
\begin{tabular}{cQc@{\hspace{4pt}}c@{}}
\toprule
& \textbf{Sub-category} & \texttt{St} & \texttt{Dy} \\ \midrule
\catcell{20}{\textsc{Specific}}
 & p4c allows generic parser/control declarations                                       & 13 & -- \\
 & p4c puts functions and types in one namespace                                        & 2 & -- \\
 & p4c internally uses \pfourinlsmall|table| to implement \pfourinlsmall|switch|        & 1 & -- \\
 & p4c expects a predefined \pfourinlsmall|NoAction| action                             & 3 & -- \\
 & p4c additionally checks parser state reachability                                    & 1 & -- \\
 & p4c additionally checks table entry reachability                                     & 1 & -- \\
 & p4c additionally checks \pfourinlsmall|packet_in| argument validity                  & 7 & -- \\
 & p4c internally converts \pfourinlsmall|struct|s to \pfourinlsmall|tuple|s on \pfourinlsmall|select| & 6 & -- \\
 & p4c performs eager constant evaluation                                               & 1 & -- \\
 & p4c treats function call expression as l-value                                       & -- & 2 \\
 & Unsupported annotations (e.g., \pfourinlsmall|@priority|)                            & 21 & 3 \\
 & V1Model (on Bmv2 switch) disallows non-byte-aligned headers                          & -- & 10 \\
 & V1Model (Bmv2) disallows writing to an \pfourinlsmall|out|-direction \pfourinlsmall|struct| & 3 & -- \\
 & V1Model (Bmv2) requires compile-time-known \pfourinlsmall|select| cases                     & 3 & -- \\
 & V1Model (Bmv2) allows non-tuples as hash arguments                                   & -- & 5 \\
 & Unimplemented \pfourinlsmall|extern_func| in V1Model tests                           & -- & 1 \\
 & Unimplemented C-function call in eBPF                                                & -- & 9 \\
 & Unimplemented \pfourinlsmall|meter| extern in PSA                                    & -- & 1 \\
 & Test expects an infinite loop                                                        & -- & 2 \\
 & Dependence on undefined value (e.g., uninitialized read)                             & -- & 13 \\ \midrule
 & \textbf{Total}                                                                       & \textbf{62} & \textbf{46} \\ \bottomrule
\end{tabular}}
\end{minipage}
\renewcommand{\arraystretch}{1.0}
\end{table}

\subsection{RQ1: Adequacy of Algorithmic Inference Rules}\label{sec:eval:algorithmic}

We evaluate the adequacy of algorithmic inference rules as a medium for
specifying the P4 language.
First, they must support the generation of a comprehensible prose specification
document.
Our generated prose specification was reviewed by the LDWG and contributed to
the decision to conditionally adopt P4-SpecTec as the official authoring
toolchain.
The generated prose specification is available in our artifact,
both in HTML and PDF format.

Algorithmic inference rules must also be expressive enough to specify the P4
static and dynamic semantics, which we assess by testing the executable
backends with tests from p4c.
Table~\ref{tab:eval:algorithmic:typechecker} and
\ref{tab:eval:algorithmic:interpreter} report the results for the P4 type
checker and interpreter backends, respectively.
Both use \texttt{-det} mode for validating determinism.
The {\small\textbf{Excluded}} column indicates the number of tests excluded
from the evaluation.
Excluded tests fall into the following categories.
Details are provided in Table~\ref{tab:eval:excluded}.

{\small
\begin{description}[style=multiline, leftmargin=6em, font=\normalfont]
  \item[\textsc{(Future)}] Features not yet standardized in P4 v1.2.5, e.g., compound assignments such as \pfourinl|+=|.
  \item[\textsc{(P4Spec)}] Under discussion with the LDWG; semantics is under-specified or inconsistent with practice.
  \item[\textsc{(P4C)}] Under discussion with p4c developers regarding non-conformance to the P4 specification.
  \item[\textsc{(Specific)}] Related to p4c-specific or target-specific implementation details, such as custom annotations.
\end{description}
}

\noindent
All {\small\textsc{P4Spec}} and {\small\textsc{P4C}} exclusions are reported to
the LDWG and p4c developers.

For the P4 type checker backend, we exclude \nTYPEExcl{} tests in total:
\nTYPEExclFuture{} categorized as {\small\textsc{Future}}, \nTYPEExclSpec{}
as {\small\textsc{P4Spec}}, \nTYPEExclComp{} as {\small\textsc{P4C}},
and \nTYPEExclSpecific{} as {\small\textsc{Specific}}.
After excluding these, the type checker backend passes all remaining
\nPOSPass{} positive and \nNEGPass{} negative tests.
We consider a negative test to pass when the type checker rejects the program,
but we do not inspect the reason for rejection.
Although not shown in the table, the P4 instantiation engine backend also
successfully instantiates all \nPOSPass{} positive tests.

Test results for the P4 interpreter backend are shown in
Table~\ref{tab:eval:algorithmic:interpreter}.
%
%
Two STF files in \VMODPlus{}, generated by P4Testgen, used commands depending
on an underspecified naming convention for referring to \pfourinl|table| keys.
We manually patched the two STF files to make the naming convention consistent
with our mechanization; these are {\small\textbf{(Patched)}} in the table.
The patches are provided in our artifact.
%
%
The {\small\textbf{Static}} column counts exclusions due to programs
already excluded from the type checker evaluation.
The {\small\textbf{Dynamic}} column counts exclusions due to dynamic semantics
issues, \nINTERPExclComp{} for {\small\textsc{P4C}} and \nINTERPExclSpecific{}
for {\small\textsc{Specific}}.
After exclusion and patching, the interpreter backend passes all tests.


\begin{table}[t]
\centering
\begin{minipage}[t]{0.45\linewidth}
\centering
\caption{P4 type checker backend results.}
\label{tab:eval:algorithmic:typechecker}
\vspace{-0.7em}
\renewcommand{\arraystretch}{0.9}
{\footnotesize
\begin{tabular}{lcccc}
\toprule
\textbf{Target} & \textbf{Pass} & \textbf{Fail} & \textbf{Excluded} & \textbf{Total} \\ \midrule
  \POSSmall{} &
  \nPOSPass{} &
  \nPOSFail{} &
  \nPOSExcl{} &
  \nPOS{}
  \\ \midrule
  \NEGSmall{} &
  \nNEGPassTbl{} &
  \nNEGFail{} &
  \nNEGExcl{} &
  \nNEGTbl{}
  \\ \bottomrule
\end{tabular}
}
\renewcommand{\arraystretch}{1.0}

\vspace{0.5em}
\begin{lstlisting}[
    keywords={add, packet, expect},
    keywordstyle=\bfseries\color{black},
    columns=fullflexible,
    basicstyle=\ttfamily\scriptsize,
    backgroundcolor=\color{blue!3},
    frame=single,
    framerule=0.1pt,
    framesep=2pt,
    rulecolor=\color{p4bg},
    xleftmargin=0.5em,
    xrightmargin=0.5em]
add access_control hdr.src:0xAA allow
packet 42 AABB
expect 1  AABB
    \end{lstlisting}
    \vspace{-1em}
    \captionof{figure}{An example STF file.}
    \label{fig:stf}
\end{minipage}\hfill
\begin{minipage}[t]{0.53\linewidth}
\centering
\caption{P4 interpreter backend results.}
\label{tab:eval:algorithmic:interpreter}
\vspace{-0.7em}
\renewcommand{\arraystretch}{0.9}
{\footnotesize
\begin{tabular}{l@{\hspace{5pt}}c@{\hspace{5pt}}c@{\hspace{5pt}}c@{\hspace{3pt}}c@{\hspace{5pt}}c}
\toprule
  \multirow{2}{*}{\textbf{Target}} & \multirow{2}{*}{\textbf{Pass (Patched)}} & \multirow{2}{*}{\textbf{Fail}} & \multicolumn{2}{c}{\textbf{Excluded}} & \multirow{2}{*}{\textbf{Total}} \\ \cmidrule(lr){4-5}
 &  &  & \textbf{Static} & \textbf{Dynamic} &  \\ \midrule
  \VMODSmall{} &
  \nVMODPassTbl{} (\nVMODPatch{}) &
  \nVMODFail{} &
  \nVMODExclStTbl{} &
  \nVMODExclDyTbl{} &
  \nVMODTbl{}
  \\
  \VMODPlusSmall{} &
  \nVMODPlusPassTbl{} (\nVMODPlusPatch{}) &
  \nVMODPlusFail{} &
  \nVMODPlusExclSt{} &
  \nVMODPlusExclDy{} &
  \nVMODPlus{}
  \\ \midrule
  \EBPFSmall{} &
  \nEBPFPassTbl{} (\nEBPFPatch{}) &
  \nEBPFFail{} &
  \nEBPFExclStTbl{} &
  \nEBPFExclDyTbl{} &
  \nEBPFTbl{}
  \\
  \EBPFPlusSmall{} &
  \nEBPFPlusPassTbl{} (\nEBPFPlusPatch{}) &
  \nEBPFPlusFail{} &
  \nEBPFPlusExclStTbl{} &
  \nEBPFPlusExclDy{} &
  \nEBPFPlusTbl{}
  \\ \midrule
  \PSASmall{} &
  \nPSAPassTbl{} (\nPSAPatch{}) &
  \nPSAFail{} &
  \nPSAExclStTbl{} &
  \nPSAExclDyTbl{} &
  \nPSATbl{}
  \\ \bottomrule
\end{tabular}
}
\renewcommand{\arraystretch}{1.0}

\end{minipage}
\vspace{-1em}
\end{table}

\subsection{RQ2: Effectiveness of Structuring}\label{sec:eval:structuring}

\begin{figure}[t]
    \centering
    \begin{minipage}[t]{0.55\textwidth}
        \vspace{0pt}
        \centering
        \captionof{table}{Comparison of EL, IL, and SL.}
        \label{tab:eval:structuring:count}
        \vspace{-0.7em}
        {\footnotesize
\renewcommand{\arraystretch}{0.9}
\begin{tabular}{lcccc}
\toprule
 & $\K{rel}$~\textbf{call} & $\K{if}$ & $\K{let}$ & \textbf{Total} \\ \midrule
\textbf{EL Prems (\#)} & \nELRulePrem{} & \nELIfPrem{} & -    & \nELPrem{} \\
\textbf{IL Prems (\#)} & \nILRulePrem{} & \nILIfPrem{} & \nILLetPrem{} & \nILPrem{} \\
\textbf{SL Instrs (\#)} & \nSLRuleInstr{} & \nSLIfInstr{} & \nSLLetInstr{} & \nSLInstr{} \\ \bottomrule
\end{tabular}
}

        \vspace{0.5em}
        \caption{Comparison of Petr4, HOL4P4, and P4-SpecTec.}
        \label{tab:eval:compare}
        \vspace{-0.7em}
        {\footnotesize
\renewcommand{\arraystretch}{0.9} 
\begin{tabular}{l@{\hspace{5pt}}c@{\hspace{5pt}}c@{\hspace{5pt}}c@{\hspace{5pt}}c}
\toprule
  \textbf{Target} &
  \textbf{Total} &
  \textbf{Petr4} &
  \textbf{HOL4P4} &
  \textbf{P4-SpecTec}
  \\ \midrule
  \POSSmall{} &
  \nPOSCompare{} &
  \nPOSPetraPass{} (\rPOSPetraPass{}) &
  - &
  \nPOSSPECTECPass{} (\rPOSSPECTECPass{})\phantom{0}
  \\
  \NEGSmall{} &
  \nNEGCompareTbl{} &
  \nNEGPetraPassTbl{} (\rNEGPetraPass{}) &
  - &
  \nNEGSPECTECPassTbl{} (\rNEGSPECTECPass{})
  \\ \midrule
  \VMODSmall{} &
  \nVMODCompareTbl{} &
  \nVMODPetraPassTbl{} (\rVMODPetraPass{}) &
  \nVMODHOLPassTbl{} (\rVMODHOLPass{}) &
  \nVMODSPECTECPassTbl{} (\rVMODSPECTECPass{})
  \\
  \VMODPlusSmall{} &
  \nVMODPlusCompare{} &
  \nVMODPlusPetraPass{} (\rVMODPlusPetraPass{}) &
  \nVMODPlusHOLPass{} (\rVMODPlusHOLPass{}) &
  \nVMODPlusSPECTECPass{} (\rVMODPlusSPECTECPass{})
  \\ \midrule
  \EBPFSmall{} &
  \nEBPFCompareTbl{} &
  \nEBPFPetraPassTbl{} (\rEBPFPetraPass{}) &
  \nEBPFHOLPassTbl{} (\rEBPFHOLPass{}) &
  \nEBPFSPECTECPassTbl{} (\rEBPFSPECTECPass{})
  \\
  \EBPFPlusSmall{} &
  \nEBPFPlusCompareTbl{} &
  \nEBPFPlusPetraPassTbl{} (\rEBPFPlusPetraPass{}) &
  \nEBPFPlusHOLPassTbl{} (\rEBPFPlusHOLPass{}) &
  \nEBPFPlusSPECTECPassTbl{} (\rEBPFPlusSPECTECPass{})
  \\ \bottomrule
\end{tabular}
}
\renewcommand{\arraystretch}{1.0} 

    \end{minipage}
    \hfill
    \begin{minipage}[t]{0.42\textwidth}
        \vspace{0pt}
        \centering
        \pgfplotsset{compat=1.18}

\begin{tikzpicture}
  \begin{groupplot}[
    group style={
      group size=1 by 2,
      vertical sep=0.5cm,
    },
    width=0.95\linewidth,
    height=3.2cm,
    symbolic x coords={IL-det,IL,SL-det,SL},
    xtick=data,
    xmin=IL-det, xmax=SL,
    ymin=0,
    enlarge x limits=0.3,
    enlarge y limits={upper, value=0.2},
    ymajorgrids=true,
    grid style={gray!30},
    axis x line*=bottom,
    axis y line*=left,
    tick align=outside,
    yticklabel style={font=\scriptsize},
    ylabel style={font=\small},
    nodes near coords={\pgfmathprintnumber[fixed,precision=0]{\pgfplotspointmeta}},
    every node near coord/.append style={font=\tiny, anchor=south},
  ]
    \nextgroupplot[
      ybar,
      bar width=10pt,
      ylabel={\POSSmall{} Time (s)},
      xticklabels={},
    ]
    \addplot[draw=black, fill=gray!55] coordinates {
      (IL-det,169.943)
      (IL,118.591)
      (SL-det,79.751)
      (SL,73.217)
    };

    \nextgroupplot[
      ybar,
      bar width=10pt,
      ylabel={\VMODPlusSmall{} Time (s)},
      xticklabel={\texttt{\small \tick}},
      xticklabel style={font=\small},
    ]
    \addplot[draw=black, fill=gray!20, postaction={pattern=north east lines}] coordinates {
      (IL-det,5558.644)
      (IL,4372.281)
      (SL-det,1004.966)
      (SL,847.521)
    };
  \end{groupplot}
\end{tikzpicture}
        \vspace{-0.5em}
        \caption{Performance comparison of IL and SL interpreters on \POS{}
for the P4 type checker and \VMODPlus{} for the P4 interpreter.}
        \label{fig:eval:structuring:perf}
    \end{minipage}
        \vspace{-1em}
\end{figure}

Structuring removes redundant premises in inference rules and inserts control
flow for efficient execution.
Table~\ref{tab:eval:structuring:count} summarizes the effect of structuring.
%
For EL and IL, we count the number of premises, where iterated premises are
counted as the premise they iterate over.
EL-to-IL introduces side conditions and $\K{let}$ premises, adding
\nELtoILDiff{} premises.
For SL, we count the number of instructions, excluding $\K{result}$, as these
do not come from premises.
Overall, \nILPrem{} premises are converted into \nSLInstr{} instructions.
In between, \nSLBranchInjected{} $\K{else}$ branches are injected.

We also compare the execution time of the IL and SL interpreters, on an Ubuntu
machine with AMD Ryzen{\texttrademark} 9 9950X 16-core CPU and 128GB of RAM.
Fig.~\ref{fig:eval:structuring:perf} shows the execution times of the IL and SL
interpreters, each with and without the \texttt{-det} mode for validating
determinism, on \POS{} (top) and \VMODPlus{} (bottom).
The results are from a single run of each interpreter.
Overall, SL is faster than IL, with a significant gap for \VMODPlus{}.
Even in \texttt{-det} mode, SL is still faster than IL.
The performance gap illustrates the effectiveness of structuring.
Furthermore, IL and SL interpreters, with and without \texttt{-det}, pass all
tests, validating semantics preservation from IL to SL.

\subsection{RQ3: Comparison with Existing Formalizations}\label{sec:eval:compare}

We compare P4-SpecTec with Petr4~\cite{petr4} and HOL4P4~\cite{hol4p4}.
P4Light~\cite{p4light} is also a P4 formalization, but is excluded because we
were unable to obtain its stable version.
Petr4 provides an OCaml implementation of P4, where we use version
v0.1.3~\cite{petr4repo}.
We use the latest version of HOL4P4~\cite{hol4p4repo} as of March 2026.
It is mechanized in Ott~\cite{ott}, with an executable P4 dynamic semantics.
However, it is unable to execute its P4 type system mechanization, internally
reusing the Petr4 type checker.

We compare the three using \POS{} and \NEG{} for P4 static semantics, and
\VMOD{} and \EBPF{} for dynamic semantics.
HOL4P4 is excluded from the type checker comparison.
\PSA{} is excluded from the interpreter comparison, because neither HOL4P4 nor
Petr4 support the architecture.
We also exclude tests mentioned in {\small\textbf{RQ1}}, as they are pending
resolution from the LDWG and p4c developers, and specific to the p4c
implementation.
%
%
We applied necessary patches to the STF files for HOL4P4, because they were not
executable without modification.

Table~\ref{tab:eval:compare} summarizes the results.
P4-SpecTec passes all \POS{} and \NEG{}, while Petr4 fails
\nPOSPetraFail{} for \POS{} and \nNEGPetraFail{} for \NEG{}.
The trend is similar for the dynamic semantics tests, where the gap is larger
for \VMODPlus{} and \EBPFPlus{}.
The two failing tests in P4-SpecTec for \VMODPlus{} also fail for Petr4 and
HOL4P4.
They are due to the issue in STF format, which we can patch to fix, as
explained in {\small\textbf{RQ1}}.
Overall, the comparison indicates that P4-SpecTec better covers P4 features
than existing formalizations.

\subsection{RQ4: Bugs Found}\label{sec:eval:bugs}

\begin{table}[t]
\centering
\caption{Specification patches made (\SP{}) and compiler bugs confirmed (\CB{}).
\texttt{St} for static and \texttt{Dy} for dynamic.
PR/Issue links to the corresponding \texttt{p4-spec} pull request (\SP{}) or
\texttt{p4c} issue (\CB{}).}
\label{tab:eval:bugs}
\vspace{-0.8em}
{\scriptsize
\renewcommand{\arraystretch}{1.20}
\setlength{\tabcolsep}{4pt}
\begin{tabularx}{\linewidth}{lcc>{\raggedright\arraybackslash}Xcc}
\toprule
& \textbf{PR/Issue} & \textbf{Section} & \textbf{Description} & \textbf{In} & \textbf{During} \\ \midrule
\textbf{SP1} & \specpr{1294} & 7.2.8 & Allow \pfourinlsmall|string| types to be aliased by \pfourinlsmall|typedef| & \texttt{St} & Mech \\
\textbf{SP2} & \specpr{1278} & 6.8 & Clarify that directionless arguments of \pfourinlsmall|extern| types are passed by reference & \texttt{St} & Mech \\
\textbf{SP3} & \specpr{1307} & 18.1 & Clarify \pfourinlsmall|const| identifiers as local compile-time known values & \texttt{St} & Mech \\
\textbf{SP4} & \specpr{1352} & 13.11 & Clarify that \pfourinlsmall|value_set|'s size is compile-time known & \texttt{St} & Mech \\
\textbf{SP5} & \specpr{1342} & 18.1 & Clarify that field, index access expressions are compile-time known & \texttt{St} & Test \\
\textbf{SP6} & \specpr{1355} & 7.2.10.2 & Clarify instantiation of extern objects with abstract methods & \texttt{St} & Test \\
\textbf{SP7} & \specpr{1341} & 6.8 & Allow implicit cast for directionless arguments in function/method/constructor calls & \texttt{St} & Test \\
\textbf{SP8} & \specpr{1356} & Appendix F & Relax runtime and compile-time call site restrictions & \texttt{St} & Test \\
\textbf{SP9} & \specpr{1343} & 7.2.8 & Add missing type nesting rules for \pfourinlsmall|extern|, \pfourinlsmall|parser|, \pfourinlsmall|control|, and \pfourinlsmall|package| types & \texttt{St} & Test \\ \midrule
\textbf{CB1} & \cbissue{5041} & - & Type inference succeeds when it should not & \texttt{St} & Mech \\
\textbf{CB2} & \cbissue{5042} & 6.8 & Directionless action argument within a P4 program should behave as direction \pfourinlsmall|in| & \texttt{St} & Test \\
\textbf{CB3} & \cbissue{5087} & 7.6 & Fail to reject implicit cast on types introduced by \pfourinlsmall|type| & \texttt{St} & Test \\
\textbf{CB4} & \cbissue{5088} & 8.11.2 & Fail to reject implicit cast from \pfourinlsmall|int<W>| and \pfourinlsmall|bit<W>| to \pfourinlsmall|int| & \texttt{St} & Test \\
\textbf{CB5} & \cbissue{5085} & 7.1.3 & Fail to reject duplicate \pfourinlsmall|match_kind| declarations & \texttt{St} & Test \\
\textbf{CB6} & \cbissue{5089} & 7.2.10.2 & Fail to reject overloading of packages & \texttt{St} & Test \\
\textbf{CB7} & \cbissue{5095} & 8.18 & Fail to reject compile-time out-of-bounds access to a header stack & \texttt{St} & Test \\
\textbf{CB8} & \cbissue{5096} & - & Fail to reject duplicate function declarations & \texttt{St} & Test \\
\textbf{CB9} & \cbissue{5256} & 7.2.8 & Fail to reject invalid type nesting & \texttt{St} & Test \\
\textbf{CB10} & \cbissue{5529} & 14.2.1.1 & Using mask (\texttt{\&\&\&}) expressions for exact match, which is incompatible & \texttt{St} & Test \\
\textbf{CB11} & \cbissue{5528} & 8.24 & Allowing \pfourinlsmall|==| comparison on variable-typed expressions, which should be disallowed & \texttt{St} & Test \\
\textbf{CB12} & \cbissue{5524} & - & Sub-parser parameters are not copied-out when transitioned to \pfourinlsmall|reject| state & \texttt{Dy} & Test \\
\textbf{CB13} & \cbissue{5521} & - & Different expects are generated for the same input packet by P4Testgen & \texttt{Dy} & Test \\
\textbf{CB14} & \cbissue{5525} & - & Tests generated by P4Testgen specify incorrect expects & \texttt{Dy} & Test \\
\textbf{CB15} & \cbissue{5539} & - & Tests generated by P4Testgen omit quotation marks for eBPF table key names & \texttt{Dy} & Test \\ \bottomrule
\end{tabularx}}
\renewcommand{\arraystretch}{1.0}

\\
\vspace{-0.1em}
\vspace{-.6em}
\end{table}

During P4 mechanization and validation of the executable backends, we
reported \nSpecIssue{} and \nCompIssue{} issues to the LDWG and p4c,
respectively.
Of these, Table~\ref{tab:eval:bugs} summarizes the \nSpecPatch{} specification
patches and \nCompBug{} confirmed p4c bugs.
\SP{1-9} denote patches we made to the official P4 specification
through discussion with the LDWG.
We clarified under-specified semantics (\SP{1-6,9}) and resolved
inconsistencies with common practice (\SP{7,8}).
\SP{7} in particular was significant, as \nSPSevenDep{} tests from
\POS{} rely on implicit casting of call arguments.
\CB{1-15} are confirmed compiler bugs.
\CB{1-11} concern static semantics, while \CB{12-15} involve dynamic
semantics.
p4c's nano-pass compiler architecture made certain type system
restrictions overlooked.
For instance, \CB{4} arises because constant evaluation pass precedes type
checking pass, allowing \pfourinl|int<W>| and \pfourinl|bit<W>| to be
implicitly cast to \pfourinl|int| during constant evaluation, contrary to
Section 8.11.2 of the P4 specification.
Testing also exposed dynamic semantics issue (\CB{12}) and issues in the
P4Testgen tool used for test generation (\CB{13-15}).

These findings highlight the value of mechanized, executable
specifications in language design.
Mechanization reveals corner cases and ambiguities in the semantics.
Testing against real programs reveals mismatches among the specification,
implementations, and even the testing infrastructure.

\subsection{Threats to Validity}\label{sec:eval:threatval}

Threats to internal validity stem from experimental setup,
where exclusion of tests may involve subjectivity.
%
%
We made certain mechanization choices to address ambiguities
in the specification, but
%
%
they are validated by the test results.
Additionally, using the p4c test suite to evaluate our P4 mechanization is not
a guarantee of its correctness.
However, the test suite includes many tests accumulated from past compiler
bugs, which we believe capture subtle corner cases in P4.
Threats to construct validity arise from using the p4c test suite for comparing
semantic coverage of P4-SpecTec, Petr4, and HOL4P4.
%
%
Nevertheless, given p4c's status as the reference compiler, the p4c test suite
serves as a valid proxy for P4 features.
Threats to external validity arise from the scope of our mechanization and
evaluation, which does not cover certain P4 features, such as
concurrency and undefined values.
However, we clearly presented our scope to the LDWG, and these limitations were
not barriers to conditional adoption of P4-SpecTec.

\section{Related Work}\label{sec:related}

\noindent
Researchers have developed various \textbf{\emph{language mechanization frameworks}}.
%
The K framework~\cite{kframework} mechanizes languages in a rewrite system,
particularly suited to modeling non-determinism and concurrency,
and derives artifacts including an interpreter and a verifier.
It has been used to mechanize specific versions of languages, including Java
1.4, C11, and JavaScript ES5~\cite{kjava, kcc, kjs}.
PLT-Redex~\cite{pltredex} is based on reduction semantics, purpose-built for
lightweight semantics engineering and testing.
%
Ott~\cite{ott} translates mechanized specifications into Rocq, Isabelle, and
HOL proof assistant code.
Skeletal Semantics~\cite{skel} defines language semantics in a Skel
meta-language, which can be used to mechanize both concrete and abstract
semantics.
The Necro framework compiles the mechanized specification in Skel into an
interpreter in OCaml and a Rocq formalization~\cite{necro}.
However, to our knowledge, none of these have mechanized
and executed the type system of a real-world language, nor
have they been integrated into real-world language standardization.

%
ESMeta~\cite{esmeta,esmetaadopt} and
Wasm-SpecTec~\cite{wasmspectec,wasmspectecadopt} are the two
\textbf{\emph{integrated mechanization frameworks}} for real-world language
design processes.
%
%
%
ESMeta compiles the prose JavaScript specification into a mechanized format,
generating a JavaScript parser, interpreter, static analyzer, and test suite
from it~\cite{jiset, jsaver, jest}.
The mechanized specification can also be type checked, discovering errors in
the JavaScript specification~\cite{jstar}.
%
%
%
Wasm-SpecTec mechanizes the Wasm specification,
%
and derives a specification document, an interpreter, and tests from
it~\cite{west}.
%
%
%
P4-SpecTec uses the SpecTec meta-language, but redesigns the backends around
algorithmic inference rules
%
to execute the mechanized P4 type system.
This is not possible with Wasm-SpecTec, as it relies on assumptions specific to
Wasm dynamic semantics for translating the declarative mechanization
into an algorithmic one.

Although not for programming languages, ASL~\cite{asl} mechanizes the ARM
architecture specification.
Similarly, Sail~\cite{sail} mechanizes instruction set architectures (ISA),
translating them to ISA simulators and proof assistant definitions.
The RISC-V Foundation adopted Sail in 2019~\cite{sailadopt}.

Ideas similar to \textbf{\emph{algorithmic inference rules}} can be found in
Centaur~\cite{centaur}, an early language mechanization framework.
Centaur uses the Typol meta-language to mechanize inference rules, and compiles
them into Prolog for execution~\cite{typol}.
It has been applied to mechanizing and executing the static semantics of Asple,
an Algol-like toy language.
Typol requires the inputs and outputs of a rule to be distinguished explicitly,
and the outputs to be deduced only when all inputs are known, resembling the
dataflow property that P4-SpecTec validates.
%
%
%
More broadly, the input/output distinction in algorithmic inference rules
relates to mode checking in logic programming, including Elf~\cite{elf}, and to
the check/synthesis distinction in bidirectional type
systems~\cite{bidirectional}.
P4-SpecTec further makes explicit the conditions for an inference rule to be
algorithmic, including determinism validation and the preclusion of unresolved
existentials.
Our novelty lies in applying algorithmic inference rules at the scale of P4 and
in generating a prose specification from them.

%
Several \textbf{\emph{P4 formalizations}} exist.
P4K~\cite{p4k} mechanizes P4$_{14}$ in the K framework, predating the P4$_{16}$
redesign.
Petr4~\cite{petr4} is the first formalization of core P4$_{16}$ using big-step
semantics.
%
It does not define the programmable parser or the static instantiation phase.
The formalization is not executable, yet the project provides a separate OCaml
implementation of P4.
%
%
%
%
%
P4Light~\cite{p4light} builds on Petr4, focusing on the instantiation phase to
verify stateful properties of instantiated P4 objects.
HOL4P4~\cite{hol4p4} mechanizes P4 in Ott using small-step semantics; while its
dynamic semantics is executable, the type system is not.
Small-step semantics enables modeling of concurrent execution,
%
but P4-SpecTec chooses big-step to provide a comprehensible specification for
P4 users.
%
Building on these, P4-SpecTec mechanizes the latest version of P4, enabling
execution of the type system, instantiation phase, and dynamic semantics.
%
%
With integration, P4-SpecTec \emph{grows with} the P4 specification, rather
than being a static snapshot of it.

%

\section{Conclusion}\label{sec:conclusion}

We presented P4-SpecTec, a language mechanization framework for P4.
From the mechanized specification, it derives a type checker,
instantiation engine, interpreter, and a prose specification.
The core of P4-SpecTec is algorithmic inference rules, which
%
%
allow direct execution of the semantics, validating the
specification with P4 programs.
%
%
Using P4-SpecTec, we identified \nBug{} bugs across the official P4
specification and the reference compiler.
%
By continued communication with the LDWG, P4-SpecTec is conditionally
adopted as the official toolchain for authoring the P4 specification.
%

The journey brings us back to the question
 raised in the introduction:
\textbf{\emph{how should mechanization be integrated into real-world languages,
lacking a normative specification?}}
%
%
Developing a mechanized specification from scratch is a daunting task; thus, we
build upon prior formalizations.
%
%
Understanding the language
 ecosystem helps identify an
appropriate mechanization medium--for P4, algorithmic inference rules.
Finally, integration is a long-term process requiring continued communication
of its values, limitations, and roadmap.
Executable specification played a key role, as the community valued
testing the specification.
%
We hope that our experience provides insights for future efforts to integrate
mechanization into real-world languages in the wild.

\begin{acks}
This work was supported by the National Research Foundation of Korea (NRF)
(2021R1A5A1021944, 2022R1A2C2003660, and 2026-25549584) and the Institute of
Information \& Communications Technology Planning \& Evaluation (IITP) grant
funded by the Korea government (MSIT) (2024-00337703).
\end{acks}

\bibliographystyle{ACM-Reference-Format}
\bibliography{ref}
\newpage
\appendix
\section{Algorithms}\label{app:algorithmic-exec}
\begin{figure}[t]
  \centering
  \begin{minipage}[t]{0.46\textwidth}
    \input{resources/algorithmic/eval/evalrel}
  \end{minipage}
  \hfill
  \begin{minipage}[t]{0.52\textwidth}
    \input{resources/algorithmic/eval/evalprem}
  \end{minipage}
\end{figure}

\textsc{Eval}$_{\textsc{Rel}}$ in Alg.~\ref{alg:evalrel} is an interpreter for
relations.
It takes the runtime environment $\Sigma$ containing the meta-definitions and
meta-values for meta-variables, a relation name $\X{id}$, and the input
$\X{val_{in}}^*$.
It first looks up $\Sigma$ for the relation definition.
Then, it performs backtracking over $\X{rule}^*$ until a successful rule
application is found (lines 3-10).
We omit the details of \textsc{Assign} and \textsc{Eval}$_{\textsc{exp(s)}}$
for brevity.
\textsc{Assign} is straightforward, as \textsc{Bind}$_{\textsc{Exp}}$ in
Alg.~\ref{alg:bindexp} has already checked that the binding expressions are
invertible and thus computable.

Premises determine whether a rule is applicable.
\textsc{Eval}$_{\textsc{Rel}}$ calls \textsc{Eval}$_{\textsc{Prems}}$ at line
7, which in turn calls \textsc{Eval}$_{\textsc{Prem}}$ in Alg.~\ref{alg:evalprem}
for each $\X{prem}$ in order.
Each case in \textsc{Eval}$_{\textsc{Prem}}$ follows the intuition, where
\K{if} premises are checked whereas \K{let} premises and relation calls bind
new $\X{id}$s and update $\Sigma$ (if successful).
The iterated premise case is omitted in the algorithm, but at a high level, it
projects the environment $\Sigma$ onto the iterated meta-variables, evaluates
the body premise, and collects the yielded meta-variables back to $\Sigma$.
For instance, suppose we are given \mbox{\specplain{(}\speckey{let}\specplain{ y = x + 1)*}} annotated \mbox{\specplain{*
}\speckeyinl{in}\specplain{ x*, }\speckeyinl{out}\specplain{ y*}} and $\Sigma$
containing \specplain{x*} $\mapsto$ \speckeyinl{[}\specplain{1,2,3}\speckeyinl{]}.
The environment is split into three copies, each containing the mapping
\specplain{x} to \specplain{1}, \specplain{2}, and \specplain{3}, respectively.
In each environment, the body is evaluated, yielding \specplain{y} mapped to
\specplain{2}, \specplain{3}, and \specplain{4}.
Finally, the projections are collected back to $\Sigma$ with \specplain{y*}
$\mapsto$ \speckeyinl{[}\specplain{2,3,4}\speckeyinl{]}.

\end{document}